\documentclass[11pt]{article}
\usepackage{epsfig}
\usepackage[usenames]{color}
\usepackage{amsmath}
\usepackage{amsfonts}
\usepackage{amssymb}
\usepackage{mathrsfs}
\renewcommand{\theequation}{\mbox{\arabic{section}.\arabic{equation}}}

\newtheorem{proposition}{Proposition}[section]
\newcommand{\bpr}{\begin{proposition}}
\newcommand{\epr}{\end{proposition}}

\newcounter{Roman}
\addtocounter{Roman}{1}

\newcommand{\beq}{\begin{equation}}
\newcommand{\eeq}{\end{equation}}
\newcommand{\bea}{\begin{eqnarray}}
\newcommand{\eea}{\end{eqnarray}}
\newcommand{\bml}{\begin{multline}}
\newcounter{saveeqn}

\newcommand{\D}{\displaystyle}

\newcommand{\ssc}{\scriptscriptstyle}

\newcommand{\re}{{\rm Re}}
\newcommand{\im}{{\rm Im}}

\newcommand{\tlS}{\tilde{S}}

\newcommand{\tllambda}{\tilde{\lambda}}

\newcommand{\tlPhi}{\tilde{\Phi}}
\newcommand{\tlphi}{\tilde{\phi}}

\input epsf
\begin{document}  

\begin{center}{\Large\bf  Nonlocal vertices and analyticity: Landau equations and general Cutkosky rule}\\[2cm] 
{Paokuan Chin\footnote{\sf e-mail: pchin@physics.ucla.edu}   and 
E. T. Tomboulis\footnote{\sf e-mail: tomboulis@physics.ucla.edu}
}\\
{\em Mani L. Bhaumik Institute for Theoretical Physics\\
Department of Physics and Astronomy, UCLA, Los Angeles, 
CA 90095-1547} 
\end{center}
\vspace{1cm}

\begin{center}{\Large\bf Abstract}\end{center}

We study the analyticity properties of amplitudes in theories with nonlocal vertices of the type occurring  
in string field theory and a wide class of nonlocal field theory models. Such vertices are given in momentum space by entire functions of rapid decay in certain (including Euclidean) directions ensuring UV finiteness but are necessarily of rapid increase in others.  
A parametric representation is obtained by integrating out the loop (Euclidean) momenta after the introduction of generalized Schwinger parameters. Either in the original or parametric representation, the well-defined resulting amplitudes are then continued in the complex space of the external momenta invariants. 
We obtain the alternative forms of the Landau equations determining the singularity surfaces showing that the nonlocal vertices serve as UV regulators but do not affect the local singularity structure. As a result the 
full set of singularities known to occur in local field theory also occurs here:  
normal and anomalous thresholds as well as acnodes, crunodes,  and cusps that may under certain circumstances appear even on the physical sheet.  
Singularities of the second type also appear as shown from the parametric representation. We obtain the general Cutkosky discontinuity rule for encircling a singularity by employing contour deformations only in the finite plane. 
The unitarity condition (optical theorem) is then discussed as a special application of the rule across normal thresholds  and the hermitian analyticity property of amplitudes.

\vfill
\pagebreak

\section{Introduction} 
\setcounter{equation}{0}
\setcounter{Roman}{0}

Nonlocal interaction vertices occur in a variety of nonlocal field theory models and in string field theory. 
In this paper we study the analyticity properties of amplitudes in theories possessing such interactions.

In contrast to the finite order polynomial momenta dependence of  local interactions, nonlocal interactions are  described by nonpolynomial functions of the momenta.  
Such vertices must satisfy certain constraints to ensure sensible physical properties. Given such vertices amplitudes can then be constructed that are well-defined as integrals along certain loop integration contours and in a certain regime of external momenta - this typically means the Euclidean regime. 
The object then is to analytically continue beyond that regime in the space of the external invariants, i.e., the various  scalar products or, equivalently, invariant energy and momentum transfer variables (Mandelstam variables) formed out of the external momenta. The analytical structure of amplitudes as a function of the external invariants viewed as complex variables is, of course, a well-studied subject in local field theory \cite{ELOP}. 
Here we examine this analytical structure in the case of nonlocal interactions.

The content of the paper is as follows. 
In section \ref{nlmods} we introduce the class of nonlocal vertices we study. 
They are constrained by various requirements including analyticity in momentum space, Lorentz invariance and UV finiteness. Requiring UV finiteness, in particular, implies that such vertices must decrease sufficiently fast (exponentially, or more generally be functions of rapid decay) in certain directions in complex  momentum space. These directions must include the Euclidean momenta so that loop integrations are well defined. 
For nonpolynomial entire functions, however, this necessarily implies exponential (or fast) growth in other directions in the complex plane, thus, generally preventing the usual Wick rotation. 
The types of vertices considered encompasses those encountered in string field theory Feynman rules and a wide class of nonlocal field models. We nominally consider such interactions within scalar field theory models but this is no real constraint - the inclusion of spin only alters vertices and propagator numerators by polynomials which cannot effect the analyticity structure in any essential way. 

Well-defined amplitudes are obtained originally in the Euclidean regime
 and are then to be analytically continued in the complex space of external invariants. It is important to note in this context that they satisfy the property of hermitian analyticity relating the matrix elements of the transfer matrix $T$ to those of its hermitian conjugate $T^\dagger$. This holds in the case of the nonlocal vertices considered here just as it does in the local case (section \ref{ampls}).  Parametric representations of amplitudes with local vertices are obtained by introducing Feynman or Schwinger parameters and integrating out the loop momenta. In the case of nonlocal vertices such explicit momenta integration after introduction of Feynman parameters is generally not possible. For a class of exponential vertices, however, such integration is possible after the introduction of (generalized) Schwinger parameters. This we do in section \ref{par}. This generalizes the parametric representation of the local theory which appears  as the zero delocalization scale limit. We also give the correspondingly generalized topological rules for computing the Symanzik polynomials of the parametric representation.

The basic tool for analyzing the singularity structure of 
amplitudes are the Landau equations \cite{L}. They locate the surfaces (algebraic varieties) in the complex space of external invariants on which singularities may reside. 
The equations are arrived at by direct examination of the integrands either in the original representation in loop momenta space; or, equivalently, in the parametric representation resulting in the parametric form of the equations. Having obtained the Landau equations in their various equivalent forms (section \ref{leqs}), one 
may proceed to examine the possible singularity structure. The central theme for us here is that any arguments or derivations that rely solely on `local' deformations of momenta or parametric contours, i.e., deformations in the finite complex plane, apply in the nonlocal case just as they do  in the local case. The parametric representation is particularly illuminating in this respect as it makes it apparent that the (appropriate) nonlocal vertices serve as UV regulators but do not modify the local structure. As a result, 
the Landau equations (section \ref{sing}) reveal the rich analyticity structure of amplitudes 
familiar from local field theory: normal and anomalous thresholds as well as other singularities such as  anodes, crunodes and Landau surface cusps. Some of the latter may, for appropriate range of the masses, appear as complex singularities even on the physical sheet. Singularities of the `second type' \cite{ELOP} need special consideration for reasons explained below. This can be properly done through the parametric representation.

The amplitude discontinuities in going around singularities in the space of invariants are given by the general Cutkosky rule \cite{C}. There are several ways of arriving at cutting rules in local field theory. For the nonlocal theories under consideration here, however, only derivations involving local use of Cauchy's theorem are possible. The original Cutkosky argument does in fact fulfill this requirement, and we adopt it to present circumstances  to derive the general rule in section \ref{cut}. This general rule applies to the discontinuity corresponding to any given solution of the Landau equations, i.e., to `cuts' across any  subset of internal lines of a given graph. 
Its particular application to normal threshold discontinuities gives the unitarity condition (optical theorem) as discussed there. 

An important feature of nonlocal interactions, however, is that knowledge of this singularity structure in the finite complex plane does not allow one to write dispersion relations of the usual  type (cf. section \ref{sing}) since one cannot close contours at infinity. This in fact appears to be {\it the} distinguishing feature of  nonlocal versus local  interactions and is related to the issue of causality. We do not study this question in this paper but comment on it in the concluding section \ref{C}. Certain technical points have been relegated to an Appendix.

There has been some related recent work on theories with nonlocal interactions of the type considered here. In \cite{PS} the discontinuity due to pinching of contours responsible for normal thresholds is computed in theories with nonlocal interactions by a careful analysis of the difference $T- T^\dagger$ and shown to be given by the Cutkosky rule. In \cite{T1} some general arguments anticipating some of the results in the present paper are given but the main focus is on the classical initial value problem with some partial consideration of the  corresponding quantum causality problem.


\section{Nonlocal field theory models \label{nlmods}} 
\setcounter{equation}{0}
\setcounter{Roman}{0}

We consider a scalar field $\phi$ with Lagrangian of the general form 
\beq 
{\cal L}(x) = {1\over 2} \partial_\mu \phi(x) \partial^\mu \phi(x) - {1\over 2}m^2 \phi^2(x) 
- \sum_nV_n[x;\phi]       \, .  
 \label{act1}  
\eeq
The interactions $V^{(n)}$ at spacetime point $x$ are taken to be functionals of the fields $\phi$ of the general form 
\beq
V_n[x; \phi]) = {\lambda_n \over n!}   \int  d^dy_1 \cdots d^d y_n \,  \hat{V}^{(n)}( x, y_1, \cdots, y_n)\, \phi(y_1) \cdots \phi(y_n)  \, .  \label{V1}
\eeq  
All couplings $\lambda_n$ are real.  
The case of local interactions, including derivative interactions, e.g., $\phi^2 \partial_\mu \phi \partial^\mu\phi$,   corresponds to 
\beq
 \hat{V}^{(n)}(x, y_1, \cdots, y_n) =  P\Big( \{\partial_y\}\Big)  \prod_{i=1}^n   \delta(x-y_i)  \,  , \label{V2} 
 \eeq
where $ P\Big( \{\partial_y\}\Big) $ is a polynomial in the $\partial_{y_i}$ derivatives; the trivial (constant) polynomial being the non-derivative $\phi^n$ interaction.   
In general, the vertices $\hat{V}^{(n)}$ are generalized functions defined as the Fourier transforms of appropriate functions: 
\beq
 \hat{V}^{(n)}( x, y_1, \cdots, y_n) = \int {d^d k_1 \over (2\pi)^d} \cdots {d^d k_n \over (2\pi)^d} \,  V^{(n)}(k_1, \cdots, k_n)    \, 
 e^{ -i[k_1(x-y_1) + \dots + k_n(x-y_n)] }             \, . \label{V3} 
 \eeq
We require that {\it the momentum-space vertices $V^{(n)}(k_1, \cdots, k_n)$ are entire functions of all their arguments}.\footnote{More specifically, entire functions of order $\geq1/2$.}  Thus, they possess no singularities anywhere in the complex $k_i$-planes (except at the point at infinity). Local interactions then comprise the subclass in which these entire functions are polynomials. 
Nonlocal  interactions result from non-polynomial entire $V^{(n)}$'s, which will be restricted by imposing 
further requirements, in particular, Lorentz invariance and perturbative UV finiteness.  In all cases reality of the vertices (\ref{V1}) implies the complex conjugation property\footnote{This  
arises from the correspondence $\partial/  \partial x_\mu \to -ik_\mu$. } 
\beq
   V^{(n)}( k_1^*, \cdots, k_n^*)    = V^{(n)}(-k_1, \cdots, -k_n)^* \, .     \label{V3a}
\eeq

The class of vertices given by 
\beq
V^{(n)}(k_1, \cdots, k_n) = \prod_{i=1}^n F(k^2_i)       \, , \label{V4}
\eeq
where $F(z)$ is entire, is of particular interest. This form of factorized vertices arises naturally as local interactions of delocalized fields. A delocalized field $\tlphi(x)$ is a functional $\tlPhi$ of the field $\phi$ with dependence on the spacetime point $x$ and generally represented  by 
\bea 
\tlphi(x) &\equiv  &  \tlPhi[x; \phi]    \nonumber \\
& = & \int d^d y \hat{F}(x,y) \phi(y)       \, . \label{V5} 
\eea
Here we take $\hat{F}$ to be a generalized function defined as the FT of an entire function $F(z)$:  
\beq 
\hat{F}(x,y) = \sum_{n=0}^\infty a_n \, \Box_x^n \,  \delta^{(d)} (x-y)   
= \int {d^d k \over (2\pi)^d}  \,  F(k^2)    \,   
 e^{-i k(x-y_1) }  
\eeq 
If now in a local scalar field potential 
\beq 
V(\phi(x)) = \sum_nV_n(\phi(x))  \geq 0    
 \,  \label{pot1} 
\eeq
one replaces $\phi(x)$ with the delocalized field $\tlphi(x)$ one arrives at (\ref{act1}) with vertices (\ref{V4}). 
Vertices of the type (\ref{V4}) posses convenient properties while incorporating all the main features of nonlocal interactions. They occur in string field theory and have been used in previous investigations of model nonlocal field theories \cite{T1}, \cite{E}.

To ensure UV finiteness we require that each $V^{(n)}$ vanishes exponentially\footnote{More generall, as functions of rapid decay.}  when one or more $k_i$'s go to infinity in certain directions in complex momentum space.  Specifically, the vertices are required to vanish in this manner along the Euclidean directions. These may be specified as usual 
by taking purely imaginary time components: $k^0_r = i \hat{k}^0_r$, with $-\infty < \hat{k}^0_r <\infty$, and real space components 
$k^j_r =  \hat{k}^j_r$, with $-\infty < \hat{k}^j_r <\infty$, and $j=1, \ldots, (d-1)$, $r=1, \ldots, n$. 
From the theory of entire functions (cf., e.g., \cite{Ti}) we know that this necessarily implies that $V^{(n)}$ will diverge (exponentially) as $k_i$'s go to infinity in other directions. This precludes the usual Wick rotation between 
Euclidean and Minkowski space.

Consider the function 
\beq 
F(k^2) = \exp [ \ell^2 k^2 ]      \, , \label{V6a}
\eeq 
where $\ell$ is some given scale. This indeed vanishes exponentially in the Euclidean directions $k^2= - [ \sum_{\mu=0}^{(d-1)} \hat{k}_\mu^2 ]$ as any $\hat{k}_\mu \to \infty$, thus rendering loop integrals UV finite.
For real Minkowski momenta, however, it vanishes exponentially only for spacelike momenta, while diverging exponentially for timelike momenta. This is unacceptable as it implies unphysical behavior in any process involving timelike exchange, e.g., in $s$-channel 2-to-2 scattering already at tree level. In complicated theories with several types of interactions, like string field theory, such unphysical behavior may be cancelled among the contributions from different vertices. For the model field theories (\ref{act1}) considered here we simply exclude such unphysical behavior by requiring that 
\beq
F(k^2) = \exp [ - {\cal P}(k^2) ]      \, , \label{V6b}
\eeq 
where ${\cal P}(k^2)$ is an even degree polynomial in $k^2$ with real coefficients and positive highest coefficient. The minimal choice is then 
\beq 
{\cal P}(k^2) = - {1\over 2} \ell_1^2 k^2 + {1\over 4} \ell_2^4 \, (k^2)^2    \,  .   \label{V6c}
\eeq
It is natural to take the coefficients proportional to some common scale: $\ell_i = c_i \ell$, $i=1,2$, where $\ell$ is the basic nonlocality scale characterizing the model. (\ref{V6c}) guarantees that the vertices asymptotically vanish exponentially inside cones around both the Euclidean and Minkowski directions. They diverge then exponentially in cones along the adjoining directions. The same holds, generally with more cone sectors, for 
(\ref{V6b}) with higher even degree ${\cal P}$.  For general vertices $V^{(n)}$ that do not necessarily factorize as in (\ref{V4}) we again take the form (\ref{V6b}) with even polynomials in variables $k$ that now stand for linear combinations of the momenta on different legs of the vertex. In all cases, one may, furthermore, include polynomials multiplying the nonlocal  $\exp [ - {\cal P}(k^2) ] $ factors. Such more general vertices encompass  the various types of vertices one encounters in string field theory (cf. \cite{PS}) and a wide class of nonlocal field theories. Apart from some technical complications, however, these further elaborations, do not, in any essential way, alter the basic analyticity structure features already present in the simpler models. 

We mention in passing that entire functions of the type considered above can be precisely characterized within the Gelfand-Shilov  theory of generalized functions. Specifically, they and their FT are naturally treated within the theory of W spaces \cite{GS}. We will not, however, need to make use of this framework for the  purposes of this paper.

The kinetic energy term in (\ref{act1}), being of the standard local form, gives the usual scalar propagator. 
The Feynman rules for (\ref{act1}) are then: \\
propagator $\D {i  \over (k^2 - m^2 +i\epsilon) }$  for each internal line of momentum $k$; and for each $n$-pronged vertex a factor:  $\D{-i {\lambda_n\over n!} V^{(n)}(k_1, \cdots, k_n)}$ with momentum conservation at each vertex. 

We have so far considered a single scalar field, but one can straightforwardly extend (\ref{act1})  to multiple scalar fields including interactions between them of the same type (\ref{V1}). This allows one to consider different masses in the various propagators and external lines without altering the basic structure of the general $L$-loop graph (cf. (\ref{A}) below) in any other way. This enhanced generality is very useful in discussing analyticity properties of amplitudes.

\section{Amplitudes and hermitian analyticity  \label{ampls}} 
\setcounter{equation}{0}
\setcounter{Roman}{0}

Consider a generic L-loop connected graph $G$ with $V$ vertices and $I$ internal lines. One has $L=I -V +1$.  
Let $P_r$ denote the total external momentum flowing into the $r$-th vertex, $r=1,\ldots, V$. 
All external momenta are taken to be incoming.  By definition, $P_r=0$ if all lines incident on the $r$-th vertex are  internal. 
By overall momentum conservation we then have 
\beq 
 \sum_{r=1}^V P_r = 0 \, . \label{momcon1}
\eeq
The internal momenta are labeled $q_j$, $j=1, \ldots, I$. Each $q_j$ is then a linear combination of the loop momenta $l_k$, $k=1, \cdots, L$, and the external momenta $P_r$. 

In the expression for the graph computed according to the above Feynman rules there is a factor of $i$ from each propagator and $(-i)$ from each vertex. 
The corresponding transfer matrix $T$ element (to the process described by graph $G$) is obtained by multiplying by an overall factor of $i$ times a total momentum conservation delta-function. Here the transfer matrix $T$ is defined 
from the $S$ matrix by $S=I - iT$.  Collecting the various factors of $i$ gives $i^{I-V+1} = i^L$. Hence, we have 
\beq
T_G = A_G \, \delta(\sum_{i=1}^V P_i)  \, ,   \label{T-A} 
\eeq
where the invariant amplitude $ A_G$ is given by 
\beq 
A_G(\{z_r\}) = C(G)\, i^L \int_{C_{\rm I}} \, \prod_{k=1}^L {d^dl_k \over (2\pi)^d}\,  \prod_{j=1}^I {1\over q_j^2 - m_j^2 +i\epsilon} \prod_{s=1}^V 
V_s^{(n_s)}(\{q,P\})   \ . \label{A}
\eeq
Here $C_{\rm I}$ denotes the loop momenta integration contours along the imaginary time-component axis  as described above. $C(G)$ denotes the product of the various coupling and graph combinatorial factors. 
The amplitude is a function of the various Lorentz invariants $z_r$ constructed from the external momenta $P_i$. For parity invariant interactions these may be taken to be the set of all scalar products $P_r\cdot P_s$, including $P_r^2$. An alternative common choice is that of total energy and all sub-energies, momentum transfers and cross-energies $z_{ijk\ldots}= (P_i + P_j + P_k + \cdots)^2$. The Mandelstam $s,t,u$ variables in 2-to-2 scattering is a familiar example. In general there are more invariants that can be formed than independent variables among them. It is best for symmetry reasons to consider the amplitude as a function of all the invariants subject to relations among them.\footnote{Note that in $d$ dimensions any set of $(d+1)$ or more vectors is linearly dependent. For an amplitude in $d$ spacetime dimensions with $E$ external legs and $E\geq d$, after imposing momentum conservation (\ref{momcon1}) and setting all $E$ external legs on-shell, there are ${1\over 2} [ 2(d-1)E - d(d+1)]$ independent variables. E.g., in 4 dimensions with $E=4$, there are 2 independent invariants, which can be any two among $s,t,u$ with one relation between the three.} 

We are interested in the properties of $A_G(\{z_r\})$ as a function of the $z_r$ promoted to complex variables. 
In the course of analytic continuation in the $z_r$ we may need to consider contours $C$ that are deformed away from $C_{\rm I}$ though still starting at $-i\infty$ and ending at $i\infty$ on each  imaginary $l^0_i$ axis. 
In  this connection the relation between the matrix elements of $T$ and $T^\dagger$ is of basic importance. 
If  $T_{fi}$ is the $T$-matrix element between an initial state $i$ and a final state $f$,  
one has $T^\dagger_{fi}= T_{if}^*$. We have the following result: 

Let the amplitude for $T_{fi}$ be given by (\ref{A}) with contour deformations $C_{\rm I} \to C$ as described above allowed.  Then the amplitude for $T^\dagger_{fi}$ is obtained by complex conjugating all external momenta and all masses (for real masses this means $(m^2 -i\epsilon) \to  (m^2 + i\epsilon)$) and changing the contour to $C^\dagger$. The contour $C^\dagger$ is obtained from $C$ by first complex conjugating $C \to C^*$ and then restoring the original orientation. Note, in particular, that $C_{\rm I}^\dagger= C_{\rm I}$.

This statement, which is easy to deduce by simple manipulations from (\ref{A}), was proven in perturbation theory in local field theory in \cite{O}\footnote{It holds also in the presence of fermions and complex couplings provided the Lagrangian is hermitian.} and the proof in the presence of nonlocal vertices, already given in \cite{PS}, is essentially the same.  It is the statement of the fundamental property of {\it hermitian analyticity} \cite{ELOP}, \cite{O}. It will be invoked in sections \ref{sing} and \ref{cut} below. 
Here we only remark that it shows $T_{fi}$ and $T^\dagger_{fi}$ to be different values of the {\it same} analytic function. 
In this manner the 
behavior of scattering amplitudes is related to the singularity structure of a single analytic function  of several complex variables  $A_G(\{z_r\})$.

\section{Parametric representation  \label{par}} 
\setcounter{equation}{0}
\setcounter{Roman}{0}
The use of Feynman or Schwinger parameters has been a very useful tool in investigating the analytic structure  of amplitudes. 
Expressing the product of propagators in (\ref{A}) by the introduction of Feynman parameters the amplitude may be written in the form 
\beq 
A_G(\{z_r\}) = C(G)\,\Gamma(I)\,  i^L \int_{C_{\rm I}} \, \prod_{k=1}^L {d^dl_k \over (2\pi)^d}\int_0^1 \prod_{i=1}^I  d\alpha_i 
\,   {\delta(1- \sum_i \alpha_i) \over [\psi(\alpha, q, P)]^I } \prod_{s=1}^V 
V_s^{(n_s)}(\{q,P\})   \, , \label{Apar1}
\eeq
where 
\beq 
\psi(\alpha,l,P) \equiv \sum_{i=1}^I \alpha_i (q_i^2 - m_i^2)    \, . \label{psi}
\eeq
This `mixed' form (\ref{Apar1}), involving integrations over both loop momenta and Feynman parameters, has proved convenient for arriving at  the Landau equations (cf. below) and discussing analyticity properties. 

In the case of local vertices, where all $V_s^{(n_s)}$ factors are polynomials in the momenta, it is possible to carry out the integrations over the loop momenta in (\ref{Apar1}) explicitly resulting in the well-known Feynman parameter  representation. 
In the case of nonlocal vertices carrying out the loop momenta integrations in (\ref{Apar1}) in closed form is, in general, no longer possible. For vertices of exponential type, however, it is natural to use Schwinger, rather than Feynman,  parameters. For a class of vertices considered above, in particular, the use of Schwinger parameters allows the loop integrations to be carried out explicitly resulting into a pure parametric representation. In the case of local vertices the Schwinger and Feynman parametric forms lead to equivalent expressions. 

To obtain the parametric form of (\ref{A}) by use of Schwinger parameters we proceed as follows. 
Let 
$\varepsilon$ denotes the $V\times I$ incidence matrix given by :
\beq
\varepsilon_{ri} = \left\{ \begin{array}{r l} 1 &   \mbox{if $q_i$ leaves vertex $r$} \\
-1 &  \mbox{if $q_i$ enters vertex $r$} \\
0 & \mbox{if $q_i$ not incident on vertex $r$}
\end{array} \right.  \label{incmatrix} 
\eeq
Momentum conservation at each vertex then implies
\beq
P_r - \sum_{j=1}^I \varepsilon_{rj} q_j = 0 \; .  \label{momcon2}
\eeq
Noting that for fixed line $j$ only two of the elements of $\varepsilon$ are non-vanishing and of opposite sign, one obtains
\beq
 \left[\sum_{r=1}^V \varepsilon_{rj} \right] = 0  \, .  \label{incmatrix1}
\eeq
(\ref{incmatrix1}) implies that the rank of $\varepsilon$ for a connected graph is $V-1$.  
Conservation of the external momenta (\ref{momcon1}) then follows from (\ref{momcon2}) 
by summing over the vertex index $r$ and using (\ref{incmatrix1}), which is in fact a necessary and sufficient condition for the compatibility of the system (\ref{momcon2}).

Since we intend to carry out the loop integrations along $C_I$ we take the time components of $l_i$ to be pure imaginary, i.e., set 
$l_i^0= i \hat{l}_i^0$; and similarly take Euclidean external momenta by setting $P_r^0= i\hat{P}_r^0$. Furthermore, use of the incidence matrix allows us to take the $I$ internal momenta $q_j$ (rather than singling out $L$ independent 
loop momenta) as integration variables subject to the constraint (\ref{momcon2}). Thus, with vertices of type (\ref{V4}), 
we can write (\ref{A}) in the form:  
\bea
&  & A_G(-\{\hat{z}_r\})\, (2\pi)^d \delta^{(d)} \Big(\sum_r \hat{P}_r\Big)   \nonumber \\
& = & C(G) (-1)^{I+L} \int \, \prod_{k=1}^I {d^d\hat{q}_k \over (2\pi)^d}\,  \prod_{j=1}^I {[F(-\hat{q}^2)]^2 \over \hat{q}_j^2  + m_j^2 } \,\prod_{r=1}^V  (2\pi)^d \delta^{(d)} \Big(\hat{P}_r - \sum_{j=1}^I \varepsilon_{rj} \hat{q}_j \Big) \, , 
 \label{A1}
\eea
with all momenta now being Euclidean. 
We introduce Schwinger parameters $\alpha_i$, $i=1, \cdots, I$, for each of the propagator factors in (\ref{A1}). 
We also introduce parameters $\beta_i$, $i=1,\cdots, I$, to represent each vertex factor of type (\ref{V6c}) in the form 
\bea 
[F(-\hat{q}^2)]^2 & = & e^{-\ell_1^2 \hat{q}^2 - {1\over 2} (\ell_2^2 \hat{q}^2)^2}    \nonumber \\
& = &  \int_{-\infty}^\infty \, \frac{d\beta}{\sqrt{2\pi}}\, e^{-\ell_1^2 \hat{q}^2 - i\beta \ell^2 \hat{q}^2 -  {1\over 2} \beta^2}  \nonumber \\
& = & \left(\frac{e}{2\pi}\right)^{1\over 2} \int_{-\infty}^\infty  \,  d\beta \, e^{-\ell_1^2 \hat{q}^2 - \ell_2^2 \hat{q}^2 
-  i\beta \ell_2^2 \hat{q}^2  +  i\beta -  {1\over 2} \beta^2}  \,. \label{beta1} 
\eea
The last line in (\ref{beta1}) is obtained by displacing the integration contour in the complex $\beta$-plane from the real axis to $(-\infty -i, +\infty -i)$. We also write   
the delta functions in (\ref{A1}) in their exponential integral representation. Thus, 
\begin{align}
&\prod_{j=1}^I {[F(-\hat{q}^2)]^2 \over q_j^2  + m_j^2 }  \,\prod_{s=1}^V  (2\pi)^d \delta^{(d)} \Big(\hat{P}_r - \sum_{j} \varepsilon_{rj} q_j \Big) \,  \nonumber   \\
 = & \left(\frac{e}{2\pi}\right)^{I\over 2} \int_{-\infty}^\infty \prod_{s=1}^V d^dx_s \int_0^\infty \prod_{k=1}^I d\alpha_k \int_{-\infty}^\infty \prod_{l=1}^I d\beta_l   \;
 \exp \left[\sum_{j=1}^I \left(- \alpha_j (\hat{q}_j^2 +m_j^2)\right) \right]   \nonumber \\
& \cdot   \exp \left[\sum_{i=1}^I \left(-(\ell_1^2 + \ell_2^2 + i \ell_2^2  \beta_i) \hat{q}_i^2    
+  i\beta_i - {1\over 2}\beta_i^2 \right) \right] 
 \exp {\left[ -i \left( \sum_{r=1}^V x_r ( \hat{P}_r - \sum_{i=1}^I \varepsilon_{ri} \hat{q}_i)\right)\right] } \,. \label{A2}
\end{align} 
Inserting (\ref{A2}) in (\ref{A1}) the $q_k$-integrations may be straightforwardly performed. To next carry out the $x_s$-integrations perform first a shift of integration variables: 
\beq 
x_s = y_s  + x_{\ssc V}\, \qquad \mbox{for} \quad s=1, \cdots, (V-1)   \, . \label{xy}
\eeq
This, as it is easily seen, has the effect that, by (\ref{incmatrix1}), $x_V$ couples only to the external momenta $P_r$. Integration over $x_{\ssc V}$ then gives a delta function factor $(2\pi)^d \delta^{(d)}(\sum_r \hat{P}_r)$ enforcing overall momentum conservation. The $(V-1)$ remaining $y_s$ integrations can then be carried out. The final result assumes the form: 
\beq
A_G(\{z_r\}) = C(G) \left({ e\over 2\pi}\right)^{I\over 2}I \!\left({1\over 4\pi}\right)^{L d/2}\!\! (-1)^{I+L} \int_0^\infty \prod_{k=1}^I d\alpha_k \int_{-\infty}^\infty \prod_{l=1}^I d\beta_l \;  
   { \exp\left[J(\alpha, \beta; P) \right] \over  \Delta(w)^{d/2}} \, . \label{Apar2}
\eeq
Here we introduced the notation    
\beq 
w_j = \alpha_j +  \bar{\ell}^{\, 2} + i\ell_2^2 \,\beta_j \, , \qquad j=1, \cdots, I \, , \label{w}
\eeq
with 
\beq 
\bar{\ell}^{\,2} \equiv \ell_1^2 + \ell_2^2   \,  .  \label{barell}
\eeq 
Then, in (\ref{Apar2}):  
\beq 
\Delta(w) \equiv \Big( \prod_{j=1}^I w_j \Big)    \det d(w)   \label{Delta}
\eeq
where $d(w)$ is the $(V-1)\times (V-1)$ matrix, of rank $(V-1)$, given by 
\beq
d(w)_{rs}    \equiv  \sum_{i=1}^I \varepsilon^{(\ssc V)}_{ri}\, {1\over w_i}\, (\varepsilon^{(\ssc V)})^T_{is}      \, .    \label{dmatrix}
\eeq
In (\ref{dmatrix}) $\varepsilon^{(\ssc V)}$ denotes the incidence matrix with the $V$-th row deleted. 
Finally, $J(\alpha, \beta; P) $  in (\ref{Apar2}) is given by 
\beq 
J(\alpha, \beta; P)  = 
Q(w; P) -\sum_{i=1}^I (\alpha_im_i^2 - i\beta_i) - {1\over 2} \sum_{i=1}^I \beta_i^2   \, , \label{J}
\eeq
where $Q(w;P)$ denotes the quadratic form:  
\beq 
Q(w;P) = \sum_{r,s = 1}^{V-1} P_r d^{-1}(w)_{rs} P_s   \, .  \label{Qform}
\eeq
(\ref{Apar2}) is written in terms of Minkowski scalar products for general (complex) momenta $P_r$, but was derived above in the Euclidean regime, i.e., for momenta $\hat{P}_r$ related by $P_r^0 = i \hat{P}_r^0$ and thus  
 $P_r\cdot P_s = - \hat{P}_r\cdot \hat{P}_s$, $z_r = - \hat{z}_r$. 

For a given diagram $G$, $\Delta(w)$ and $Q(w; P)$ may be computed directly from their definitions (\ref{Delta}) and (\ref{Qform}), respectively. Alternatively, the expressions for them can be conveniently obtained from the topology of the graph. To give a precise statement of  these topological prescriptions we need a few basic graph theory notions. Consider the Feynman diagram $G$ as a graph defined  by $V$ vertices and $I$ edges (the internal lines).\footnote{External lines of $G$ are irrelevant for the graph-theoretic considerations here.}   A  tree in $G$ is a subgraph that contains no loops (circuits, in common graph theory terminology). The number of vertices $V_T$ and edges $I_T$ of tree $T$ are related by $V_T = I_T +1$. 
A spanning tree $T$ in $G$ is a subgraph that is a tree and contains all the vertices of $G$, i.e., $V_T=V$. Note that then $I - I_T = L$. The complement $T^*$ of a spanning tree $T$ in $G$ is called a co-tree (or the `chord set' corresponding to $T$). Thus, $I_{T^*} = I - I_T = L$. 
A $2$-tree $T^{(2)}$ is a spanning tree with one edge removed. Removing one edge of  a spanning tree leaves two vertex-disjoint connected subgraphs $T^{(2)}_{\pm}$ making up the 2-tree: $T^{(2)} = T^{(2)}_{+} \cup T^{(2)}_{-}$. 
Note that, from its definition, a  2-tree contains all the vertices of $G$ and that one of the subgraphs $T^{(2)}_{\pm}$ may be an isolated vertex.\footnote{More generally,  
an $n$-tree $T^{(n)}$ is a spanning tree with $(n-1)$ edges removed. Thus a spanning tree is a 1-tree. An $n$-tree contains all vertices of $G$  and consists of $n$ connected components.}  
The compliment of a $2$-tree in $G$ is the corresponding 2-co-tree $T^{(2)*}$. 

We can now give a precise statement of the topological rules for obtaining $\Delta(w)$ and $Q(w; P)$. 
The determinant of the matrix $d(w)$ defined in (\ref{dmatrix}) is given by 
\beq
\det d(w)  =   \sum_{T \subset G} \, \prod_{i\in T} {1\over w_i}    \, .  \label{dmatrix1}
\eeq
It follows from (\ref{Delta}) that 
\beq 
\Delta(w) =   \sum_{T \subset G} \, \prod_{i\in T^* }  w_i     \, . \label{Deltatop} 
\eeq 
As seen from (\ref{Deltatop}) $\Delta(w)$ is  a homogeneous polynomial in the $w_i$'s of degree $L$. 
$Q(w;P)$ is then obtained from 
\beq 
\Delta(w) \,Q(w; P) =   \sum_{T^{(2)} \subset G} \, z_{T^{(2)}} \prod_{i\in T^{(2)*} } w_i   \; .    \label{discr1}
\eeq
The variables $z_{T^{(2)}}$ denote the square of the sum of the momenta $P$ entering either of 
$T^{(2)}_{\pm}$:
\beq 
z_{T^{(2)}} =\left( \sum_{r\in T^{(2)}_+} P_r \right)^2 =  \left( \sum_{r\in T^{(2)}_-} P_r \right)^2     \, . \label{discr2}
\eeq
The r.h.s. of (\ref{discr1}) is a homogeneous polynomial in the  $w_i$'s of degree $L+1$. Thus $Q(w;P)$ is given by the ratio of two polynomials and is homogeneous of degree $1$. 
The relation (\ref{dmatrix1}) is well-known from the theory of circuits. The proof of (\ref{discr1}), following  \cite{Na}, is more involved.\footnote{The quantities $\Delta(a)$ and $\Delta(\alpha) Q(\alpha,P)$ are known as Symanzik polynomials. Here they are generalized to complex parameters $w$.} 

Furthermore, one has the important fact that 
 \beq
 |\Delta (w) | \not= 0 , \quad \mbox{for} \quad \bar{\ell}^2 \not=0   \, \label{Delta1}
 \eeq
for all $\alpha_i \geq 0$, $-\infty < \beta_i < \infty$. We prove (\ref{Delta1}) in the Appendix.

The local case is obtained by setting $\ell_1=\ell_2=0$: 
\beq
A_G(\{z_r\}) = C(G)\left({1\over 4\pi}\right)^{L d/2}\!\! (-1)^{I+L} \int_0^\infty \prod_{k=1}^I d\alpha_k 
   { \exp\left[J(\alpha, 0; P) \right] \over  \Delta(\alpha)^{d/2}} \, . \label{Apar2a}
\eeq
Rescaling $\alpha_i \to \rho\alpha_i$ in (\ref{Apar2a}) and integrating over $\rho \in (0,\infty)$ gives the same result that is obtained from (\ref{Apar1}), with $V_s=1$ vertex factors, by integrating over the loop momenta. Explicitly, the result is given by (\ref{Is1}) below with $s=0$ and $\rho_0=0$ (so only the $k=0$ term survives on the r.h.s. of (\ref{Is1})). These manipulations are valid provided (\ref{Apar2a}) and the local (\ref{Apar1}) version are UV convergent, which will be the case for $(I-\frac{d}{2}L)>0$. Otherwise,  as it well known, there is UV divergence from the $\alpha_i=0$ Schwinger parameter integration endpoint where $\Delta(\alpha)=0$, as evident from 
(\ref{Deltatop}) with $w_i=\alpha_i$. In the local case then, (\ref{Delta1}) can be violated. This divergence can be regulated by cutting off the low end of the Schwinger parameter integration range. We now see explicitly from the representation (\ref{Apar2}) that this precisely what the nonlocal vertices accomplish: {\it they render (\ref{Apar2}) UV finite by the introduction of the nonlocality scale $\ell$.} This effects the shift $\alpha_i \to \alpha_i + \bar{l}^2$, cf. (\ref{w}), thus ensuring (\ref{Delta1}) (cf. Appendix). {\it Otherwise, the analyticity structure of the amplitude in the finite complex plane remains unaffected} as it is seen by examining the Landau equations for it, which we do next.

\section{Landau equations and singularity structure}
\setcounter{equation}{0}
\setcounter{Roman}{0} 

Let us briefly recall the general procedure whose application leads to the Landau equations. 
Consider a function $I(z)$ of a set of complex variables $z=\{z_r\}$, $r=1,2,\ldots$, given by an integral representation 
\beq
I(z) = \int_C \, dw f(w,z)      \label{irep1}
\eeq
for some contour $C$  in the space of the complex integration variables $w=\{w_s\}$, $s=1,2,\ldots$, and some domain $R$ in $z$-space for which the integral (\ref{irep1}) is well-defined. One may then extend the definition of $I(z)$ outside $R$ by analytic continuation. 
A general method for doing this goes back to \cite{H}, and was 
elaborated in the physics literature in \cite{PS1}, \cite{PS2}; for a mathematically more rigorous homology-based approach see \cite{HT}. 
As $z$ moves outside of the region $R$, the singularities $\{w_s(z)\}_i$, $i=1, \ldots$, of the integrand in (\ref{irep1}) move in the complex $w$-space and may approach the contour C,  
which may then need to be deformed to avoid them.  
This will cease to be possible either if two or more of these singularities pinch the contour; 
or if a singularity hits the fixed boundary of  C. 
A general formulation encompassing these possibilities is given as follows \cite{PS1}, \cite{PS2}.  Let the location of singularities of the integrand be given by a set of equations $S_i(w,z)=0$, $i=1,2, \ldots, n$; and let the boundaries of the contour be specified by another set $\tlS_ j=0$, $j=1,2,\ldots,m$. Introduce corresponding parameters $\lambda_i$, $i=1,\ldots,n$, and $\tllambda_j$, $j=1,\ldots,m$. The critical hypersurfaces on which the singularities of $I(z)$ are located are then specified by the solutions of the following conditions \cite{PS1}, \cite{PS2}: 
\begin{enumerate}
\item[(i)]
\beq 
\lambda_i S_i(w,z) =0 \, , \qquad \mbox{for each} \ i  \; ; \label{i} 
\eeq
\item[(ii)]
\beq  
  \tllambda_j \tlS_j(w,z) =0 \, , \qquad \mbox{for each} \ j  \; ; \label{ii}
 \eeq
\item[(iii)]
\beq  
 {\partial \over \partial w_k} \Big(\sum_i \lambda_iS_i(w,z) + \sum_j \tllambda_j \tlS_j(w,z) \Big) = 0 \, \qquad 
 \mbox{for each} \ k   \; .   \label{iii} 
   \eeq
\end{enumerate}
The nice feature of the method is that one has only to examine the integrand for its singularities in complex space and apply these conditions. Working out their implications, however, can be highly nontrivial. 

Application of conditions (i) - (iii) to the Feynman integral representation of an amplitude $A(\{z_r\})$ gives the Landau equations \cite{L}, also \cite{Bj}. Depending on which form of the Feynman integral one uses these equations are given in different equivalent forms.

\subsection{The Landau equations \label{leqs}} 

The first form of the Landau equations is obtained from the representation (\ref{A}) in which 
there are no integration boundaries. With all vertex factors $V^{(n_s)}$ given by entire functions, the only singularities in the integrand can only come from the denominators: $S_i= (q_i^2 - m_i^2) $. 
Hence, application of (\ref{i}) gives 
\beq 
\lambda_i (q_i^2 - m_i^2) = 0 \, , \qquad \mbox{each} \ i   \; , \label {L1a} 
\eeq 
i.e., 
\beq 
\mbox{either} \quad q_i^2 = m_i^2    \, , \qquad \mbox{or} \quad \lambda_i=0   \qquad \mbox{for each} \ i \, , \label{L1b}
\eeq
where $i= 1, \ldots, I$ enumerates the internal lines. From (\ref{iii}) and noting that  
each internal momentum $q_i$ is a linear combination of the loop momenta and the external momenta one  obtains   
\beq 
\sum_{i\in C_k} \lambda_i q_i=0 \;,  \qquad \mbox{for each} \ k  \, . \label{L1d}
\eeq
In (\ref{L1d}) the sum runs around the loop $C_k$, where $k=1,\cdots, L$ enumerates a set $\{C_1, \cdots, C_L\}$ of $L$ independent loops in the graph. 

(\ref{L1b}) and (\ref{L1d}) are the Landau equations for the graph $G$ (in what is often referred to as the first representation). The important point for us here is that they are independent of the form of the vertices as long, of course, as the vertices are entire functions: polynomials in the local case and transcendental functions (such as (\ref{V6b})) in the nonlocal case. In either case, they, being entire, introduce no other singular points in the integrand and thus lead to the same form (\ref{L1b}) and (\ref{L1d}) of the Landau equations. 
Strictly speaking we have so far considered only the finite complex plane. In the extended complex plane singularities may occur at the point at infinity, which then require special consideration. These are in fact the so-called singularities of the second type. We will return to them below. 

According to (\ref{L1b}) and (\ref{L1d}) singularities arise from configurations where either an internal line is on shell or the corresponding parameter $\lambda_i=0$. For a given graph the solution with all $\lambda_i\not= 0$ gives its leading singularity. Those with some of the $\lambda_i= 0$ are the non-leading (or lower order) singularities. A non-leading singularity is the leading singularity of the corresponding so-called reduced graph, i.e., the graph obtained from the given graph by contracting each internal line with $\lambda_i= 0$ to a point.

One may alternatively use the `mixed' form (\ref{Apar1}). 
To discuss analyticity properties using the representation (\ref{Apar1}) first note that the $\alpha$-integration can be extended to $+\infty$ since the delta-function enforces $\alpha_i\leq 1$. We then multiply (\ref{Apar1}) by 
$1= \int_0^\infty d\rho \,e^{-\rho}$  and rescale $\alpha_i \to  \alpha_i/\rho $. Carrying out the $\rho$ integration (\ref{Apar1}) is recast in the form
\bml
A_G(\{z_r\}) = \\ C(G)\,\Gamma(I)\,  i^L \int_{C_{\rm I}} \, \prod_{k=1}^L {d^dl_k \over (2\pi)^d}\int_0^\infty \prod_{i=1}^I  d\alpha_i \,  
 {(\sum_{j=1}^I \alpha_j) \, e^{- (\sum_{j=1}^I \alpha_j) } \over [\psi(\alpha, q, P)]^I }  
 \prod_{s=1}^V  V_s^{(n_s)}(\{q,P\}) \, . \qquad \label{Apar1a}
\end{multline}
With entire vertices there is now only one singularity surface $S=\psi(\alpha,q,P)$, and boundary surfaces 
$\tlS_i=\alpha_i$. Straight application of conditions (i) - (iii) then gives a set of equations which are easily reduced to 
\bea 
\alpha_i (q_i^2 - m_i^2) & = &  0 \, ,  \qquad \mbox{each} \ i = 1, \cdots, I   \label{L2e}\\
\sum_{i\in C_j} \alpha_i q_i  & = & 0 \;,  \qquad \mbox{each} \ j= 1, \cdots, L \, ,  \; , \label{L2f}
\eea
These are again the Landau equations (\ref{L1a}) and (\ref{L1d}) (with the relabeling $\lambda_i = \alpha_i$). For future reference we note that (\ref{L2e}) - (\ref{L2f}) can also be expressed as: 
\beq 
\psi =  0 \, ,  \quad \mbox{and } \quad   {\partial \psi \over \partial l_j}= 0  \qquad \mbox{each} \ j \label{L2g}
\eeq 
and 
\beq 
\alpha_i {\partial \psi \over \partial \alpha_i} = 0
 \qquad \mbox{each} \ i  \; . \label{L2h}
\eeq

The parametric form of the Landau equations, after having performed the momentum integrations, is obtained from (\ref{Apar2}).   Let $\zeta_i \equiv \bar{\ell}^2 + i\bar{\beta}_i$, with $\bar{\beta}_i \equiv \ell_2^2 \beta_i$, so that $w_i = \alpha_i + \zeta_i$, and also $Q_{, i \ldots j}(w)  \equiv {\D 
\partial Q(w) \over  \D \partial w_i \ldots \partial w_j }$. 
Now, 
\beq
Q(w) = 
Q(\alpha) + \sum_{i=1}^I \zeta_i Q_{,i}(\alpha) +  \sum_{i, j=1}^I \zeta_i \zeta_j \int_0^1 ds \int_0^s dt \, Q_{,ij}(\alpha+ t\zeta)     \, . \label{Qexp1}
\eeq
Since $Q(w)$ is a homogeneous function of degree $1$, $Q_{,i}$ and $Q_{,ij}$ are homogeneous of degree $0$ and $-1$, respectively. 
To examine the large $\beta$ regime, consider the rescaling $\beta_i \to \rho \beta_i$. From (\ref{Qexp1}) one has 
\bml
Q(\alpha+ \bar{\ell}^2 +i\bar{\beta}) =   Q(\alpha) +  \bar{\ell}^2 \sum_{i=1}^I Q_{,i}(\alpha) 
+ i 2\bar{\ell}^2   \sum_{i, j=1}^I \bar{\beta}_i \int_0^1 ds \int_0^s dt \, Q_{,ij}({\alpha+  \bar{\ell}^2 \over \rho} +i t\bar{\beta})  \\ 
 + \rho\left[ i \sum_{i=1}^I \bar{\beta_i}Q_{,i}(\alpha)   - 
  \sum_{i, j=1}^I \bar{\beta}_i \bar{\beta}_j \int_0^1 ds \int_0^s dt \,Q_{,ij}({\alpha+  \bar{\ell}^2 \over \rho} +i t\bar{\beta}) \right]         \, . \label{Qexp2}
\end{multline}
Hence, $Q(w)$ grows at most linearly with $\rho$, plus $O({\rm constant})$ and $O({1/ \rho})$ corrections.  
The large $\beta$ behavior of $J(\alpha,\beta; P)$ is then always controlled by the $-{1\over 2}(\sum_i \beta_i^2)$ term, thus resulting in convergent behavior for $\beta_i \to \pm \infty$. 

On the other hand,  from (\ref{Qexp1}), under $\alpha_i \to \rho\alpha_i$ one has
\beq
Q(\rho\alpha + \zeta) =   
\rho Q(\alpha)  + \sum_{i=1}^I \zeta_i Q_{,i}(\alpha) + {1\over \rho}  \sum_{i, j=1}^I \zeta_i \zeta_j \int_0^1 ds \int_0^s dt \, Q_{,ij}(\alpha+ t\zeta/\rho)  \,   \label{Qexp3}
\eeq 
and, hence,  for $\rho \to \infty$: 
\beq 
J(\alpha,\beta; P) =  \rho \left[Q(\alpha; P) - \sum_{i=1}^I \alpha_im_i^2 \right]   + O({\rm constant}) + O({1\over \rho})  
  \label{J1}
 \eeq
 For large $\alpha_i$ then convergence requires that 
 \beq 
 S \equiv    Q(\alpha; P) - \sum_{i=1}^I \alpha_im_i^2    \label{S1}
 \eeq
satisfies $S<0$. This is the case for Euclidean momenta, and $\alpha_i \geq 0$, for which the parametric representation was derived. The identifies S=0 as a singularity surface. 

Note that the region of both $\alpha$ and $\beta$  simultaneously  large, i.e., $\alpha_i \to \rho\alpha_i$,  $\beta_i \to \rho\beta_i$, $\rho \to \infty$, gives nothing new - it is equivalent to the large $\beta$ regime. 

We may in fact  single out the contribution of non-vanishing $\alpha$ parameters in (\ref{Apar2}) explicitly by inserting a delta function in the integrand. Omitting numerical factors this contribution is given by 
\bml
I =  \int_0^\infty \prod_{k=1}^I d\alpha_k \int_{-\infty}^\infty \prod_{l=1}^I d\beta_l  \int_{\rho_0}^\infty  d\rho \, 
 \delta(\rho - \sum_{i=1}^I \alpha_i) \\
 {1\over \Delta(w)^{d/2}}\,  \exp\left[ Q(w; P) -\sum_{i=1}^I (\alpha_im_i^2 - i\beta_i) - {1\over 2} \sum_{i=1}^I \beta_i^2 \right] \, ,   \label{Apar3}
\end{multline}
with $\rho_0 >  0$. The delta function constraints one or more  $\alpha_i$ parameters to be nonzero, the actual value  of $\rho_0$ being irrelevant for this purpose.  Rescaling $\alpha_i \to \rho \alpha_i$, using (\ref{Qexp3})  together with 
\beq
\Delta(\rho\alpha + \bar{\ell}^2 + i\ell_2^2\beta) = \rho^L \Delta(\alpha + { \bar{\ell}^2 + i\ell_2^2\beta \over \rho}) \, , \label{Deltaexp1}
\eeq  
and expanding in powers of $1/\rho$, one obtains a series of integrals over $\rho$ of the form 
 \beq 
I_s \equiv  \int_{\rho_0}^\infty d\rho \rho^{[I-{d\over 2}L -1]} \left({1\over \rho}\right)^s \exp \left[ - \rho \left[-Q(\alpha;P) + \sum_{i=1}^I \alpha_i m_i^2\right]\right]   \,  \label{Apar4}
 \eeq
with integer $s\geq 0$. Carrying out the  $\rho$ integration for Euclidean momenta, for which $Q(\alpha;P) <  0$, i.e., $S< 0$,  and 
with\footnote{Even $d$ is assumed so $q$ is integer.}  
\beq 
q\equiv  I-{d\over 2} L -s   \, , \label{q}
\eeq 
one obtains:  
\begin{align} 
I_s =&  e^{ \rho_0 S(\alpha;P) } \sum_{k=0}^{q-1} {\Gamma(q)\over k! } \frac{\rho_0^k}{[-S(\alpha; P)]^{(q-k)}}  \, , \quad   &  \mbox{for} \quad q > 0   \;  ;  \label{Is1} \\ 
   \nonumber     \\
I_s = & -  {\rm Ei}(\rho_0 S(\alpha;P))  \, ,  \qquad &  \mbox{for} \quad q = 0  \; ;  \label{Is2}  \\  
     \nonumber     \\
I_s =  & { (-1)^{|q| +1} \over \Gamma( |q|+1)}  [-S(\alpha; P)]^{|q|}\, {\rm Ei}(\rho_0 S(\alpha;P))   \quad  &  \nonumber  \\
 &  +   
{e^{ \rho_0 S(\alpha;P) } \over \rho_0^{|q|} } \sum_{k=0}^{|q|-1} {(-1)^k \rho_0^k\,  [-S(\alpha;P)]^k \over |q| ( |q| - 1) \cdots (|q|-k) }   \, , 
\quad   & \mbox{for} \quad  q <  0  \; ,  \label{Is3}
\end{align}  
where 
\beq
{\rm Ei}(x) =   {\bf C} + \ln (\pm x) + \sum_{k=1}^\infty \frac{x^k}{k k!}  \, , \qquad x \gtrless 0  \label{Ei}
\eeq  
is the exponential integral function with $\bf C$ denoting Euler's constant. As seen directly from (\ref{Is1}), (\ref{Is2}) then, upon continuation  
$S=0$ is indeed the  singularity surface. $I_s$ is regular for sufficiently large $s$   as evident from (\ref{Is3}). 

Note that the $s=0$ contribution is the exact result for the local theory obtained by setting $\ell_1=\ell_2=0$, i.e., $V_s^{(n_s)}=1$, 
in the Schwinger parametric representation (\ref{Apar2}). As pointed out right after (\ref{Apar2a}) this is identical to the result obtained by carrying out the momentum integrations in the Feynman parameter representation (\ref{Apar1}), provided the integral is UV convergent; 
otherwise it will have to be UV regulated. In the nonlocal case the transcendental entire vertices automatically provide such regularization, whereas the singularity structure defined by the  $S$ surface remains intact. 

Applying the conditions (i) - (iii), with, again, $\alpha_i=0$ defining the contour boundaries, we 
now have:\footnote{We disregard the trivial $\lambda=0$ case in the $\lambda S(\alpha,P)= 0$ equation.}
\bea 
S(\alpha,P) & =& 0   \label{L3a} \\ 
\tllambda_i \alpha_i & = & 0  \label{L3b} \\ 
{\partial \over \partial \alpha_i} \Big( \lambda S+ \sum_{k=1}^I \tllambda_k \alpha_k\Big) & = & 0 \, , \qquad \mbox{each} \ i   \; .  
 \label{L3c} 
 \eea
Hence, either $\tllambda_i=0$ and $\partial S/\partial \alpha_i=0$, or  $\alpha_i=0$. But 
\beq
S= \sum_{i=1}^I \alpha_i {\partial S\over \partial \alpha_i} \, ,   \label{EulerS}
\eeq
since $S$ is homogeneous of degree one in the $\alpha_i$. Thus one concludes that 
\begin{align} 
 \mbox{either:} &  \qquad \alpha_i=0     \label{L3d} \\ 
 \mbox{or:} & \qquad {\partial S\over \partial \alpha_i} = 0   \; ,  \qquad \mbox{each } \quad i     \, .   \label{L3e}  
 \end{align} 
In either case then (\ref{L3a}) is automatically satisfied.  For many purposes the parametric form of the Landau equations (\ref{L3d}) - (\ref{L3e}) turns out to be the most convenient form for analyzing the analyticity properties of amplitudes. 

The parametric form of the Landau equations (\ref{L3d}) - (\ref{L3e}) is equivalent to that given by  (\ref{L2g}) - (\ref{L2h}). This is shown in the Appendix. Note that this implies that the former are in fact of general validity 
 even though they were obtained within the vertex subclass used in deriving the parametric representation.  

\subsection{Singularity structure  \label{sing}}  

Eliminating the $\lambda$ or $\alpha$  parameters from the Landau equations one obtains the equations for the Landau surfaces (algebraic varieties) in the multi-dimensional space of the external invariants $z_r$. 
This can be done by forming the inner product of (\ref{L1d}), or, equivalently, (\ref{L2f}), with $q_j$ giving a set of simultaneous equations for the $\lambda$, or, respectively, $\alpha$, parameters. The  condition for non-trivial solution of this set gives the Landau surface equations. Equivalently, they may be derived from the condition for non-trivial solution of the system of equations given by (\ref{L3e}). In this manner the equations for the leading Landau surface $\Sigma(z_r)$ may be given in the form 
\beq 
\det Y= 0  \, ,    \label{Lvar1}
\eeq
where $Y(z_r)$ is a   
matrix with entries depending on the external momenta invariants $z_r$. The lower order surfaces  $\Sigma_{ ij\cdots}(z_r)$, corresponding to solutions with a subset of the $\alpha_i$'s (resp., $\lambda$'s) equal to zero, are then given by the vanishing of the appropriate minors of $Y$. 

\begin{figure}[ht]
\begin{center}
\includegraphics[width=12cm]{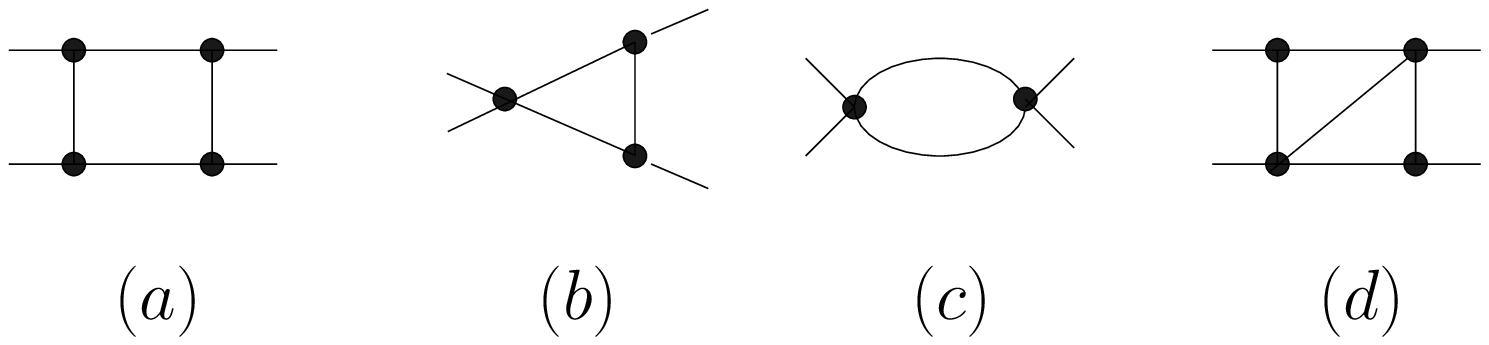}
\end{center}
\caption{(a) The one-loop box graph; (b) one of the once-reduced graphs of (a) containing the anomalous thresholds; (c) one of the twice-reduced graphs of (a) containing the normal thresholds; (d) the acnode graph. Black circles denote nonlocal vertices.    \label{F5.1}}
\end{figure}

Thus, for example,  for the basic square box diagram in Fig. \ref{F5.1}(a), with external masses $M_i$, internal masses $m_i$ and external legs on shell, there are two independent invariants, the familiar  $s$ and $t$ Mandelstam variables. 
Proceeding as indicated above one straightforwardly finds the leading singularity  Landau surface $\Sigma$ given by 
(\ref{Lvar1}), 
where 
\beq 
Y_{ij}(z_r) = - { z_{ij} - m_i^2 - m_j^2     \over 2 m_i m_j}    \, , \quad i\not= j \, ; \qquad Y_{ii} = 1 \, . \label{Lvar2}
\eeq
Here $z_{ij}=( \sum_{r \subset ij} P_r )^2$ denotes the square of the sum of the external momenta entering between lines $i$ and $j$. Thus, $z_{13} = (P_1+ P_4)^2 = t$, $z_{24} = (P_1 + P_2)^2 = s$, whereas $z_{ij}= z_{ji} = P_j^2= M_j^2$ for adjacent $i$ and $j$ with $i< j$ in cyclic order clockwise around the graph. 
The invariants formed by the squares of the external momenta have been set on shell; considering them as variables amounts then to continuation in the dependence on the external masses. 
The surfaces for the subleading singularities are found from the corresponding reduced graphs. Setting one $\alpha$ 
parameter to zero gives the triangle graphs in Fig. \ref{F5.1}(b), whose leading singularities are the `anomalous' thresholds. Each of their Landau surfaces $\Sigma_i$, $i=1,2,3,4$, may be obtained from the corresponding reduced graph and, as easily seen, is given by 
\beq 
\det Y[i,i] = 0  \, , \label{Lvar3}
\eeq
where $Y[i,i]$ denotes the submatrix of $Y$ obtained by deleting the $i$-th row and $i$-th column of $Y$; i.e., $\Sigma_i$ is given by the vanishing of the $(i,i)$-minor of $Y$.  

Similarly, solutions with two $\alpha$ 
parameters equal to zero correspond to 
Landau surfaces $\Sigma_{ij}$ given by 
\beq
\det Y[ij,ij] = 0  \, , \label{Lvar4}
\eeq 
where $Y[ij,ij]$ denotes the submatrix obtained by deleting the $i$-th and $j$-th rows and the $i$-th and $j$-th columns of $Y$. In the case of the square graph these are the leading singularities of the reduced graphs in Fig. \ref{F5.1}(c), which include the usual normal thresholds, and (\ref{Lvar4}) results, respectively, in:
\beq 
s = (m_2 \pm m_4)^2   \, , \qquad \mbox{and}  \qquad   t= (m_1 \pm m_3)^2   \; . \label{Lvar5}
\eeq

Landau surfaces generally have rather intricate structure with multiple components in the multi-dimensional space of the external invariants. This is true even for the simplest diagrams, such as the square graph and the reduced triangle graphs in Fig. \ref{F5.1}(b)\footnote{The triangle  
may of course be viewed in its own right as a form factor depending on general complex $z_i=P_i^2$, i.e., a 6-dimensional space.}  given by (\ref{Lvar2}). Such  examples have been extensively studied, cf. \cite{ELOP} for a review.  
One typically starts with the solution for the real section of the surface in the complex space of the invariants, e.g., the intersection of $\Sigma$ with the real $(s,t)$ plane in the example of the box diagram above; this section will, in general, consists of several components.  The complex parts attached to the real sections are then determined by the searchline technique \cite{Tar}. This can be a rather laborious procedure. Having determined the shape of the surfaces on which potential singularities lie, there remains the nontrivial task of determining which parts are actually singular on which Riemann sheet. 
In the present context, however, the main point, once more, is that the resulting local singularity structure is not affected by replacing polynomial by transcendental entire vertices, since, as we saw, except for providing UV regularization, this leaves the Landau equations unaffected. 

It will be useful for our discussion to recall here some aspects of this singularity structure. 
Despite the generally complicated form of the Landau surfaces 
some general features are present that are crucial for the structure of physical amplitudes.  The physical sheet is defined by attaching a small imaginary part $-i\epsilon$ to each internal mass  $m_i^2$ and integrating over undistorted integration contours in (\ref{Apar1}) or, equivalently, (\ref{Apar2}). Now, from (\ref{discr1}),  
\beq 
\im \,S(\alpha, P) = {1\over \Delta(\alpha)}  \sum_{T^{(2)} \subset G} \, \im \,z_{T^{(2)}} \prod_{i\in T^{(2)*} } \alpha_i + (\sum_{i=1}^I \alpha_i m_i^2 ) \,\epsilon   \;  .\label{imS1} 
\eeq
For real positive $\alpha$'s (undistorted contours), this does not vanish for $\im\, z_r\,  \geq 0$. The same, of course, holds for the imaginary part of the extremized form (\ref{psi}) (w.r.t. the loop momenta) which equals $S$, see Appendix. Hence, one may continue through this region of complex $z_r$ space without the 
potential singularity $S = 0$ or $\psi=0$ in the integrands forcing a distortion of the contour. The physical amplitude is obtained in the limit $\im \, z_r \to 0^+$. This defines the direction of approach to the subspace of real $z_r$ in which one may encounter singularities. We know, however, that there is a region $R$ in real $z_r$ subspace, which includes Euclidean momenta,  where $S\not= 0 $. The boundaries of this region will be determined by  those parts of the real sections of the Landau surfaces $\Sigma, \Sigma_i, \Sigma_{ij}, \ldots$ that correspond to real, positive $\alpha$ solutions of the Landau equations. The common situation is that the normal thresholds provide the boundaries of this analyticity region $R$; at the  boundary provided by the first threshold for two particle intermediate state the amplitude acquires a cut, followed by additional cuts at the onset of three and higher intermediate states. This is the familiar singularity structure expected from unitarity. The general situation, however, can be rather more complicated due to the existence of the anomalous thresholds and other, nastier types of singularities. These, upon continuation in the $z_r$, may move to replace the normal thresholds as boundaries.

The 1-loop box diagram of Fig. \ref{F5.1}(a) provides an instructive example. One has 
\beq 
S(\alpha; s,t)= {\alpha_1 \alpha_2 M_2^2 +   \alpha_1\alpha_4 M_1^2 + \alpha_1\alpha_3\, t + \alpha_2\alpha_4 \,s
+\alpha_2\alpha_3M_3^2 + \alpha_3\alpha_4 M_4^2 
\over \alpha_1 + \alpha_2 + \alpha_3 +\alpha_4}   - \sum_{i=1}^4 \alpha_i m_i^2   \, .   \label{Sbox}
\eeq
One chooses the external masses to be stable against decay into pairs of the internal masses, i.e., $M_1 < m_1 + m_4$, etc. Then for sufficiently small masses\footnote{More precisely, for masses such that none of the $\Sigma_i$ surfaces corresponds to a solution of the non-leading triangle Landau equations with real positive $\alpha$. 
The effect of subsequent increases in the masses violating this condition successively for each $\Sigma_i$ is described in the main text. } one can show that $S<0$  for $s < (m_2 + m_4)^2$ and $t < (m_1 + m_3)^2$. Thus the boundaries of the region $R$ in this case are given by the normal threshold surfaces $\Sigma_{13}, \Sigma_{24}$. This, however, changes drastically as  the masses are increased \cite{ELPTa}. As the external masses are increased one of the triangle surfaces, say $\Sigma_2$, moves to collide with the normal threshold surface $\Sigma_{24}$ and then separate. As a result the boundary of $R$ now consists of $\Sigma_{13}$ and $\Sigma_2$ and an anomalous threshold singularity appears on the physical sheet. Increasing the external masses further another $\Sigma_i$ moves up to and then away from $\Sigma_{13}$ and the region $R$ is now bounded by two anomalous threshold surfaces $\Sigma_i$. Further increases of the masses leads to collision of the $\Sigma_i$'s with the leading surface $\Sigma$. The $R$ boundary is now formed by part of $\Sigma$ and the $\Sigma_i$  and this is accompanied by the onset  of complex  singularities on the physical sheet \cite{ELPTa}.  This disturbing appearance of physical sheet complex singularities originates here in the movement of the anomalous thresholds as function of the masses.  

Unfortunately, this is not the only known mechanism by which complex singularities can appear on the physical sheet. Investigation of the 2-loop box diagram shown in Fig. \ref{F5.1}(d), the infamous acnode graph, reveals a new phenomenon \cite{ELPTb}, \cite{OT}. In addition to the continuous parts of the real section of surface $\Sigma$, as some external mass is increased beyond a certain value isolated points, known as acnodes, appear on the real $(s,t)$ plane. There are singular complex pieces of the surface, of complicated shape, intersecting the real plane only through these acnodes. The acnodes move as the masses are further increased till they meet the continuous part of the real surface section creating a crunode (self-intersection point) as well as cusps in $\Sigma$. The physical interpretation of such physical sheet complex singularities and their effect on physical amplitudes remains 
obscure.\footnote{ Their appearance is what prevents the general validity of the Mandelstam representation in the case of local (polynomial) vertices \cite{ELPTb}.}

Whatever the boundaries of the real region $R$ of analyticity may be, 
its existence has some important consequences.  Analytic continuation in $z_r$-space  out of $R$ along a path $P$ to some point $z_r$ is related to that along the complex-conjugate path $P^*$ to point $z_r^*$ by complex conjugation. Thus, any distortion of integration contours made necessary during the continuation along $P$ implies the complex-conjugate distortion along $P^*$. If, therefore, analyticity can be proven in one region, it is guaranteed to hold also in the complex conjugate region. Furthermore, a path out of $R$ with $\im \, z_r >0$ to an endpoint with $\im\, z_r = \epsilon^+$, i.e., on the physical sheet, and the complex conjugate path to the conjugate point relate the amplitude for the transfer matrix $T$ to that for $T^\dagger$ by continuation of the {\it same} analytic function. This is the content of the hermitian analyticity property of section \ref{ampls} above. The introduction of nonlocal entire function vertices does not in any way interfere with these fundamental properties.  

The existence of the real analyticity region $R$, in fact, generally implies extension to a larger region of analyticity \cite{Wu}. First note that 
since $S< 0$ with undistorted contours for $z_r \in R$, $S$ cannot vanish if one continues to complex $z_r$ such that $\re \,z_r \in R$. Thus the amplitude is analytic in a tube in the complex $z_r$-space whose real section is $R$. Furthermore, given a fixed point $z_r^0 \in R$, consider the line in complex $z_r$-space given by $z_r(\zeta) = z^0_r + \lambda_r \zeta$ with complex variable $\zeta$ and given numbers $\lambda_r$ defining its direction. Now, since by (\ref{discr1})  $S$ is linear in $z_r$, for points on the line one has 
\beq 
S(\alpha; \zeta) = S(\alpha; z_r^0) + {\cal T}(\alpha) \,\zeta   \, ,   \label{L1} 
\eeq 
where ${\cal T}$ is a function of only the $\alpha$. For $\alpha$'s real and positive, i.e., on the undistorted contours, one then has 
\beq 
 \re\, S(\alpha;\zeta) = S(\alpha; z_r^0) + {\cal T}(\alpha) \re \,\zeta  \; , \qquad   \im \, S(\alpha;\zeta) = {\cal T}(\alpha)\im \,  \zeta  \; .  \label{L2}
\eeq
It follows from this that $S(\alpha;\zeta)$ cannot vanish when $\im \, \zeta \not= 0$. Indeed, for it to vanish $\im \, S= 0$, implying that ${\cal T}=0$; and $\re\, S=0$, implying then that $S(\alpha; z_r^0)=0$, which is not true, since $z_r^0 \in R$. Hence, the amplitude is analytic for all $\zeta$ such that $\im\, \zeta \not= 0$ and all real values of $\zeta=x$ such that that $z_r(x) \in R$. Thus, if, for example, the boundaries of $R$ are given by the normal thresholds, the amplitude is analytic in the entire $\zeta$ plane except for the normal threshold cuts along the real axis. 

The  result is independent of the type, polynomial or nonlocal entire functions, of the vertices.
It immediately allows one to obtain single-variable dispersion relations {\it provided} contours in complex $z_r$-space can be closed at 
infinity.\footnote{For example, for 2-to-2  scattering, where $z_r= (s,t)$, one can choose the line through the region $R$ in the above argument such that one obtains, say,  a dispersion relation in $s$ at fixed $t$.}   This, modulo possible subtractions, is possible for polynomial interactions,  but clearly not for nonlocal interactions. So, even though the singularity structure is the same in the finite complex $z_r$-space, the different behavior at the point at infinity in the nonlocal case does not allow one to use the above result to write such dispersion relations  in any obvious way. For both classes of vertices, however, one may, of course, employ Cauchy's theorem by closing contours in the finite plane, as in fact is done in arriving at the Cutkosky discontinuity rule (cf. below). 

Finally, there is the matter of the so-called second-type singularities \cite{FLNP}, \cite{C}. 
As originally discovered in the representation (\ref{A}) with conventional polynomial interactions,  these are singularities that appear to correspond to pinches of the loop integration contours at infinity. To make such discussions mathematically more well defined one has to make change of variables to convert these to finite points. A better discussion, which, most importantly for us here, can be equally well applied to nonlocal entire as well polynomial vertices, is given in the parametric representation.  
Second-type singularities can then be defined as solutions to the Landau equations (\ref{L3d}) - (\ref{L3e}), or, equivalently, (\ref{L2g}) - (\ref{L2h}), which {\it additionally} satisfy $\Delta(\alpha)=0$.  Now we proved (see Appendix) that $\Delta(\alpha)\not=0$ for all $\alpha_i \geq 0$. (In the local case, this holds only for $\alpha_i > 0$.) In fact, $\Delta(\alpha)$ could vanish only for some $\alpha_i < -\bar{\ell}^{\, 2}$ due to the explicit UV regularization provided by the nonlocal vertices. Second-type singularities occur then due to pinches with distorted contours of negative $\alpha$. In all known examples these negative $\alpha$ values imply that the singularity is not on the physical sheet. The general situation regarding the Riemann sheet properties of second-type singularities, however, is an open 
question.

\section{Cutkosky discontinuity rule  \label{cut}} 
\setcounter{equation}{0}
\setcounter{Roman}{0} 

Consider the amplitude (\ref{A}) and let $\{z_r\}= \{z_r^0\}$ denote a singularity given by a solution of the Landau equation with $I_c$ lines on-shell and $(I-I_c)$ of the $\alpha$ parameters equal to zero. The Cutkosky rule provides a general formula for the discontinuity from going around the singularity $z^0$.

Let $L_c$ be the number of independent loops in the reduced diagram obtained by contracting the $(I-I_c)$ lines with zero $\alpha$. We can label lines and choose the loop momenta so that the reduced graph has internal line momenta $q_i$, $i=1, \ldots, I_c$ and they depend only on the loop momenta $l_j$ with $j=1,\ldots, L_c$.   
Let us label loop momenta components by $l_{j\mu} \equiv l_A$ with $A$ assuming $dL_c$ values, and define the $I_c \times dL_c$  Jacobian matrix 
\beq 
J_{iA} = {\partial q_i^2 \over \partial l_A} \, .  \label {jac1}
\eeq
Assume that $I_c \leq dL_c$ and that the rank of $J$ equals $I_c$.\footnote{The argument may be extended to cases where these assumptions are violated, but there is no need to consider such cases in this paper.}  Then, following \cite{C}, one may introduce a change of variables  replacing the loop momenta $l_A$ on the contour $C_I$, i.e., Euclidean loop momenta, by the set $\xi_A$, where $\xi_i= \hat{q}_i^2=-q^2_i$, $i=1, \ldots, I_c$, and additional variables $\xi_a= \chi_a$, $a= 1, \ldots, (dL_c - I_c)$. The new variables may be visualized geometrically in $d$-dimensional Euclidean space as the construction of a simplex whose base is the polygon, closed by momentum conservation, formed by the vector diagram of the external momenta vectors.
The rest of the oriented edges represent the $I_c$ internal momenta vectors. 
The simplex on top of this base is built 
in  such configuration of edges that momentum conservation is obeyed at each vertex of the reduced diagram. 
The variables $\xi_i = \hat{q}^2_i$ then represent the squared lengths of the edges and the set $\chi_a$ represent the additional `angle' variables needed to completely specify, for given $\hat{q}_i^2$,  the allowed distortion/orientation degrees of freedom of the simplex in the $dL_c$-dimensional  space. It may be that this latter set can be chosen in more than one way. 
This construction is always possible in Euclidean space showing that, under the above assumptions, this is a well-defined change of variables in (\ref{A}).  

The experts will recognize this as the first step in the construction of the so-called dual diagram for the graphical solution of the Landau equations.\footnote{Also introduced in  
\cite{L}. See \cite{ELOP} for references and examples.} Were one to additionally impose the Landau equations they would result into further  relations between the edges that will, in general, not be possible to satisfy in Euclidean space and thus require complex or Minkowski vectors. This, of course, reflects the occurrence of singularities outside the original region of definition of the integral, which are encountered upon continuation in the external invariants. 

In terms of the new variables (\ref{A}) assumes the form
\bml 
A_G(\{z_r\}) = {C(G)\,(-1)^{(L_c+ I_c)}  i^{(L-L_c)} \over (2\pi)^{dL}}\, \prod_{i=1}^{I_c}  \int_{a_i}^{b_i}dq^2_i   \int \prod_{a=1}^{(dL-I_c)}
  d\chi_a \; \int \prod_{j > L_c}^L dl_j \, {1\over |J|} \\
 \cdot \,  \prod_{k=1}^I
    {1\over (q_k^2 - m_k^2)} \prod_{s=1}^V V_s^{(n_s)}(\{q^2, \chi, l,P\}) \, .  \label{C1}
\end{multline}
We choose to express (\ref{C1}) in terms of $q^2_i$ rather than $\hat{q}^2_i= -q^2_i$ since we eventually want to consider complex $q^2_i$ contour deformations as the external invariants move out of the Euclidean region - cf. remark in the previous paragraph. 
Also, the masses $m_k^2$ are normally real with small negative imaginary parts, but, more generally, may be allowed to assume any complex values. 
The limits of integration $a_i, b_i$ are determined by the extrema of $q^2_i$ w.r.t.  the reduced graph loop momenta for fixed $q^2_k$, $k< i$, i.e., by  extremizing 
\beq
\phi(l, \gamma) = q^2_i + \sum_{k<i} \gamma_k q^2_k     \,  \label{lmult1}
\eeq 
w.r.t. to $l_j$ ($1\leq l_j \leq L_c$) and  Langrange multipliers $\gamma$. 
Note that one may always choose to label internal lines with $q_i= l_i$ for $i=1, \cdots L_c$, so that for these lines one simply has $\hat{q}^2_i = \hat{l}_i^2$ and, hence, $(a_i, b_i) =( -\infty, 0)$.  (We necessarily have $1\leq L_c < I_c$). Subsequent $a_i, b_i$ limits, however, will generally depend on the preceding $q^2$'s and the $z_r$'s. 
\begin{figure}[h]
\begin{center}
\includegraphics[width=15cm]{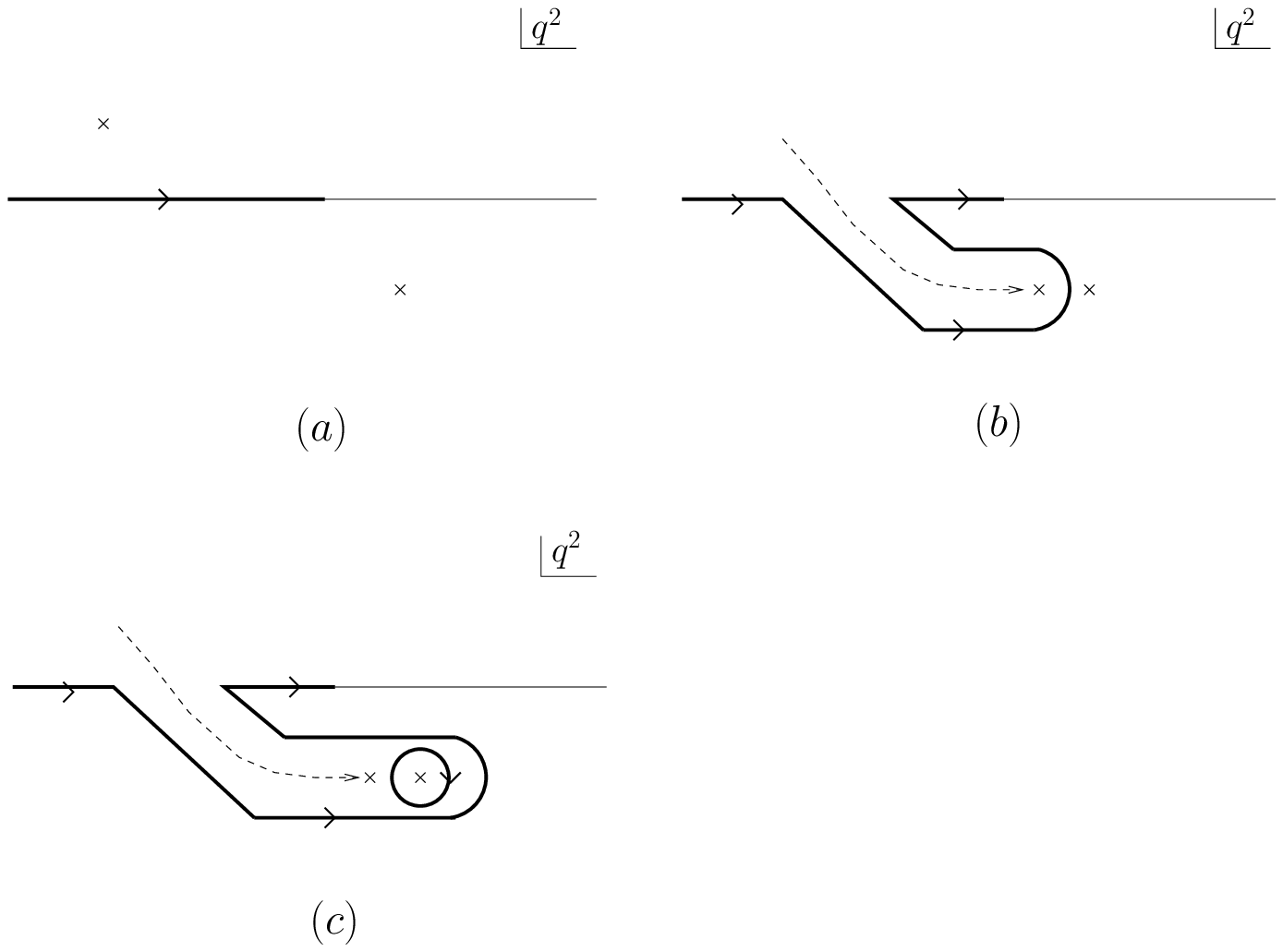}
\end{center}
\caption{(a) Original position of singularities with undistorted contour in $q^2$ plane; (b) singularity movement as $z\to z^0$ pinching the contour; (c) equivalent contour consisting of  regular and a singular (circle) contribution.    \label{F6.1}}
\end{figure} 

Now write (\ref{C1}) in the form
\beq
A_G(\{z_r\}) = K   \int_{a_1}^{b_1} dq^2_1  {1\over (q_1^2 - m_1^2)}   I^{(1)}(q^2_1)   \, , \label{C2}
\eeq
where $K\equiv {C(G)\,(-1)^{(L_c+ I_c)}i^{(L-L_c)} /  (2\pi)^{dL}}$. The integrand 
$I^{(1)}(q^2_1)$ is the result of having integrated over all variables in (\ref{C1}) except $q^2_1$ and singling out the $q^2_1$-propagator factor explicitly - any corresponding $q^2_1$-dependent vertex factors (cf. structures as in (\ref{V4})) have been lumped in the definition of $I^{(1)}(q^2_1)$.  Note that, by the remark above, one can always have $a_1= -\infty$, $b_1=0$.

By assumption, as $z\to z^0$ (\ref{C1}) has a singularity involving all $I_c$ lines of the reduced graph being on-shell. The only way that this can happen in (\ref{C2}) then is if the pole at $q^2_1= m_1^2$ is pinched by a singularity in $I^{(1)}(q^2_1)$ as $z\to z^0$ (Fig. \ref{F6.1}(b)). The distorted contour may be moved through the pole and split into two pieces: one on which the integral is regular and a small circle around the pole (Fig. \ref{F6.1}(c)).
The integral over the former, being regular, will not contribute to the discontinuity. The latter will be pinched and is the one that is singular. By the residue theorem it is given by 
\beq 
K (-2\pi i) \, I^{(1)}(m_1^2) =K  (-2\pi i)  \int_{a_2}^{b_2} dq^2_2  {1\over (q_2^2 - m_2^2)}   I^{(2)}(m_1^2, q^2_2)   \, , \label{C3}
\eeq
where  $I^{(2)}(q^2_1, q^2_2)$ is the result of having integrated over all variables in (\ref{C1}) except $q^2_1$ and $q^2_2$ and after singling out the $q^2_2$- and $q^2_1$-propagator factors explicitly - (and , again, having lumped  any corresponding vertex factors in  $I^{(2)}(q^2_1, q^2_2)$).  

The same argument can now be applied to $I^{(2)}(m^2_1, q^2_2)$. Iterating the argument $(I_c-1)$ times, the singular part is found to be given by the expression 
\beq
K(-2\pi i)^{(I_c-1)}  \int_{a_{I_c}}^{b_{I_c}} dq^2_{I_c}  {1\over (q_{I_c}^2 - m_{I_c}^2)}  I^{(I_c)} (m_1^2, m^2_2, \cdots, m_{(I_c-1)}^2, q^2_{I_c}) \,.  \label{C4}
\eeq
The limits of integration are here given by the extrema of (\ref{lmult1}) with $i=I_c$. In this case, however, as it is easily seen,\footnote{By rescaling of the $\alpha$'s one may always set one of them equal to one. The Lagrange multipliers $\gamma$ are here  the parameters $\alpha$.} the extremizing equations are identical to the Landau equations (\ref{L2e}) and (\ref{L2f})  for the leading singularity of the reduced graph, i.e., by assumption, the singularity $z^0$. Hence, in (\ref{C4}) the singularity is the result not of contour pinching but of an endpoint singularity: either $b_{I_c}$  or $a_{I_c}$ moves toward $q^2_{I_c}= m^2_{I_c}$ as $z\to z^0$ thus giving rise to  an endpoint singularity.  
\begin{figure}[ht]
\begin{center}
\includegraphics[width=15cm]{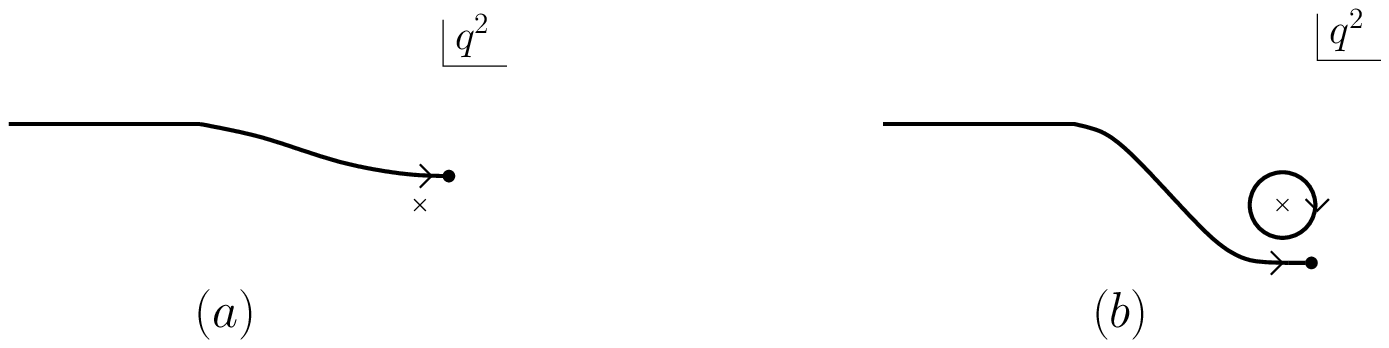}
\end{center}
\caption{Difference (circular contour) between continuing: (a) above and (b) below endpoint singularity.     \label{F6.2}}
\end{figure} 
The discontinuity from going around the endpoint singularity in (\ref{C4}) is then obtained as depicted in Fig. \ref{F6.2} and given by 
\beq
\Delta A_G(\{z_r\}) = K(-2\pi i )^{I_c} I^{(I_c)}(m_1^2, \cdots, m_{I_c}^2)   \, .  \label{C5}
\eeq
There are several remarks to be made concerning (\ref{C5}). 
\begin{enumerate}
\item  The above derivation, which harks back to the original Cutkosky argument, is well suited for extension in the presence of nonlocal vertices of the general type introduced in section \ref{nlmods}. 
The change of variables is well-defines in Euclidean space and the subsequent argument involves only local deformations of contours, i.e., in the finite complex plane, as the external invariants $z_r$ are continued. This is in contrast to other ways of arriving at cutting-rules such as employing \cite{T2} the Feynman \cite{F} or tree-loop theorems \cite{tloop}, which involve closing contours at infinity; or space-time techniques such as the largest-time equation \cite{V} which are not suited to handle nonlocal interactions. 
\item
The usual case is that of real masses with small negative imaginary parts, $m^2_i - i\epsilon$, corresponding to physical region (real momenta) singularities. In fact the figures above are drawn depicting this type of relative positioning of contours and singularities. In particular, all signs from evaluation of residues confirm to this situation.
In this case then, reverting to the original variables (\ref{C5}) may be expressed in the form:
\bml 
\Delta A_G(\{z_r\}) \\
= C(G) i^L
 \int_{C}  \prod_{k=1}^L {d^dl_k \over (2\pi)^d}  \prod_{j=1}^{I_c}(-2\pi i) \delta^+(q_j^2 - m_j^2)  \prod_{i= I_c + 1}^I 
 {1\over q_i^2 - m_i^2} \prod_{s=1}^V 
V_s^{(n_s)}(\{q,P\})  \, . \label{C6}  
\end{multline}
This is the Cutkosky discontinuity rule as usually stated. The contours $C$ in (\ref{C6}) are specified as follows. The  integration over loop momenta that circulate in the reduced graph may be taken over Minkowski space since the delta functions set these internal lines on-shell - they correspond to the little circular contours in figures 2, 3. The loop integrations in the parts of the graph that would be contracted in the reduced graph must be along $C_{\rm I}$, i.e., Euclidean directions. The momenta external to these parts, which  are either external lines or reduced graph internal lines,  are then to be continued to Minkowski space. This pertains to point 4. below.

\item In the case of singularities at complex momenta (\ref{C6}) should be viewed merely as a mnemonic device for (\ref{C5}) since the delta functions in (\ref{C5}) are not immediately defined for complex arguments. The signs of some of the $(2\pi i)$ factors in (\ref{C5}) may need to change in this general situation. As we saw above such singularities may in certain circumstances occur even on the physical sheet. 
Explicit evaluation of the discontinuity in these cases may not be easy in practice. 

\item A given singularity in the amplitude $A$ for a given process is shared by all contributing Feynman graphs that can be contracted to the corresponding reduced graph. Each graph contributes to the discontinuity according to (\ref{C6}) so that the sum gives the complete discontinuity in the form 
\beq 
\Delta A(\{z_r\}) = 
 i^{L_c}  \int_{C}  \prod_{k=1}^{L_c} {d^dl_k \over (2\pi)^d}  \prod_{j=1}^{I_c} (-2\pi i) \delta^+(q_j^2 - m_j^2)  \prod_{i=1}^{V_c}  A_i(q,P)
\, , \label{C7}
\eeq
where $V_c$ is the number of vertices in the reduced graph, and the $A_i$'s are the amplitudes represented by 
\begin{figure}[ht]
\begin{center}
\includegraphics[width=9cm]{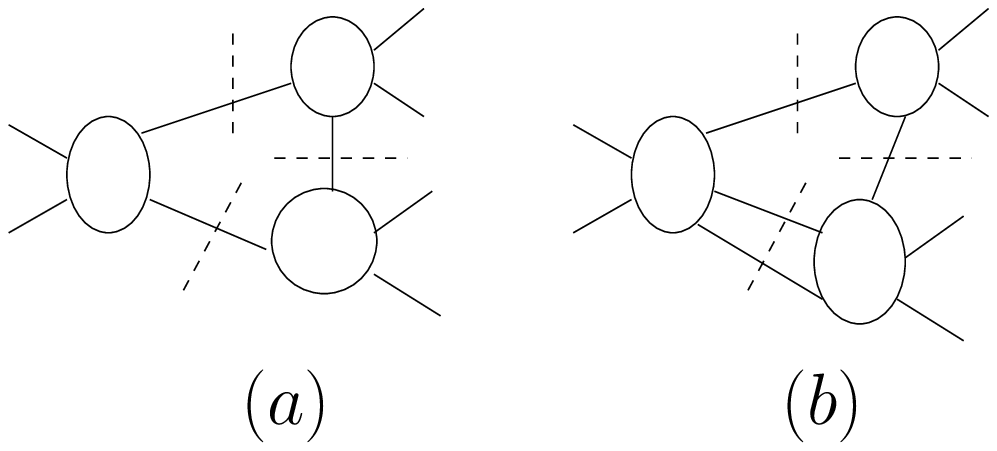}
\end{center}
\caption{Examples of eq. (\ref{C7}) for discontinuities involving: (a) three (anomalous threshold); and (b) four Cutkosky cuts (on-shell lines).  \label{F6.3}}
\end{figure}
the blobs in the examples in Fig. \ref{F6.3}. The following issue may now arise. (\ref{C7}) gives the discontinuity obtained by going around the singularity $z^0$ in question, which is the leading singularity of the reduced graph. In doing so, however, it is not immediately clear what Riemann sheet each $A_i$ may end up on; this will depend on whether or not one is simultaneously going around some singularity in $A_i$. To decide this requires a separate analysis in each particular case. 
\end{enumerate}

As a particular application consider the normal threshold singularity in a given channel $s$. It is specified as the Landau equations solution for the leading singularity of the (set of) reduced graph(s) such that a cut through the internal lines separates the graph(s) into two pieces along the direction of the given channel. 
Application of (\ref{C7}) then gives an expression for the discontinuity across this singularity. 
\begin{figure}[ht]
\begin{center}
\includegraphics[width=12cm]{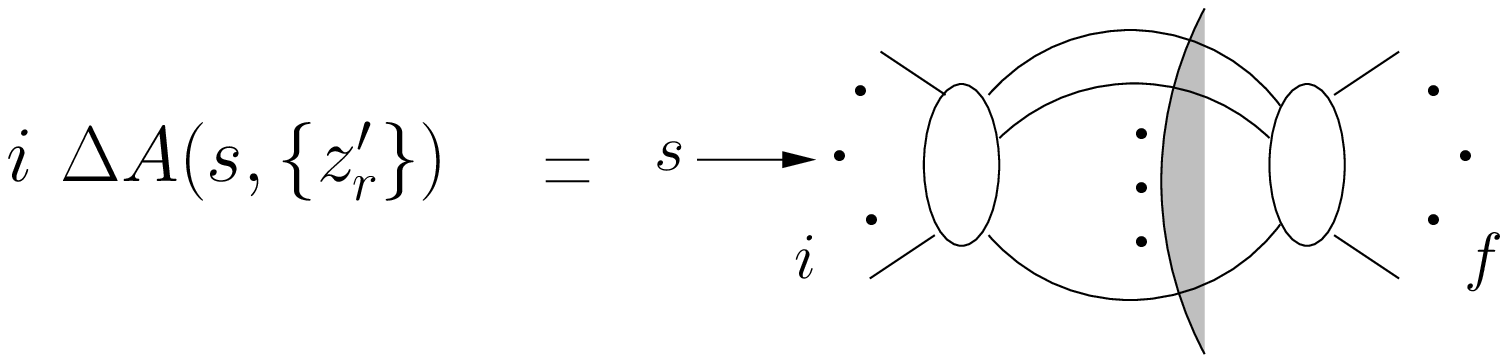}
\end{center}
\caption{Discontinuity across normal threshold given by eq. (\ref{C8}). Energy flows from unshaded to shaded side of cut (on-shell) lines. Feynman rules on shaded side are those for $T^\dagger$.    \label{F6.4}}
\end{figure}
Noting  that $i^{L_c}(-i)^{I_c} = i^{(-V_c+1)} = -i$ since here $V_c=2$, (\ref{C7}) gives (figure 5): 
\beq 
\Delta A(s, \{z^\prime_r\})  = -i \int_{C}  \prod_{k=1}^{n-1} {d^dl_k \over (2\pi)^d}\;  [A_{fn}]\left( \prod_{j=1}^{n} (-2\pi ) \delta^+(q_j^2 - m_j^2) \right)   A_{ni}   \, . 
\label{C8}  
\eeq
In (\ref{C8}) we placed the outgoing  $A_{fn}$ `blob' in square brackets to indicate that, as pointed out above, the 
general Cutkosky rule does not a-priori specify its sheet placement ($i\epsilon$ prescription).  
The total discontinuity in $s$ for 
the amplitude $A_{fi}$ between states $i$ and $f$ is given by summing over all possible intermediate state discontinuities (\ref{C8}) for the channel in question:  
\beq
 A_{fi}(s+i\epsilon, \{z^\prime_r\}) - A_{fi}(s-i\epsilon, \{z^\prime_r\})  =  
\sum  \Delta A(s, \{z^\prime_r\}) 
\, . \label{C9}
\eeq
So far we have simply applied the Cutkosky rule to the normal threshold discontinuity across a given channel $s$. This discontinuity is, of course, known to be related to the optical theorem. We may arrive at this connection by the following indirect argument. 
The l.h.s. of (\ref{C9}) represents the difference between the amplitude where the real section in the space of the invariants is approached from above and that where it is approached from below. For the normal threshold singularity under consideration only the invariant $s$ is discontinuous by definition. The  property of hermitian analyticity (section \ref{ampls}) applied to all the contributing reduced graphs implies then 
\beq 
A_{fi}(s+i\epsilon, \{z^\prime_r\}) - A_{fi}(s-i\epsilon, \{z^\prime_r\})  =    A_{fi}(s, \{z^\prime_r\}) - A^\dagger_{fi}(s,\{z^\prime_r\})  \, .  \label{C10}
\eeq
Complex conjugating (\ref{C10}) shows that the only consistent choice for the factor in square brackets in (\ref{C8}) is $[A_{fn}] = A_{fn}^\dagger$, which results in the familiar form of the optical theorem. 
Note, however, that this is not a first principles derivation of unitarity. What the argument actually shows is that the normal threshold discontinuity as given by the Cutkosky rule combined with the hermitian analyticity of the contributing reduced graphs gives a statement of a form consistent with the optical theorem. Nonetheless, this type of  indirect consistency argument invoking hermitian analyticity can be very useful. Thus, for example, specializing to the case of only 2-particle intermediate states it can be extended to conclude that these normal thresholds  singularities can only be two-sheeted \cite{BGMT}.\footnote{$n$-particle normal threshold singularities with $n\geq 3$ are conjectured to be infinite-sheeted but no general argument is apparently known, cf. \cite{ELOP}.}

To reiterate, for us the important point here is that, having obtained the general Cutkosky rule in the presence of  nonlocal interactions, any of its various applications carry through as long as only local contour deformations are involved in the argument.

\section{Concluding remarks \label{C}}

In this paper we examined the analyticity properties of amplitudes under continuation in the external invariants in theories with nonlocal interactions. 
We found that appropriately chosen nonlocal interactions serve as UV regulators but leave 
the analytical structure locally, i.e., in the finite complex `energy plane' (complex space of external invariants) intact. This, in particular, is made manifest in the parametric representation obtained by integrating out the loop momenta after the introduction of generalized Schwinger parameters. As a result the Landau equations yield the full familiar complement of singularities, i.e., the expected physical region normal and anomalous thresholds, as well as those resulting from the movement and intersection of anomalous thresholds, acnodes, Landau surface cusps, etc. Some of these more exotic singularities may, under certain circumstances (cf. section \ref{sing}), appear even on the physical sheet. Singularities of the second type also appear in the same manner as can be seen from the parametric representation. The general Cutkosky rule (section \ref{cut}) gives the discontinuity upon encircling a given singularity. 

If anything, this discussion of the local analyticity structure is made more rigorous in the nonlocal case since amplitudes are manifestly UV finite. The local case is obtained as the limit where the delocalization scale characterizing the nonlocal vertices is taken to zero - one has then to include the contribution of  subtraction terms if any are needed. 

There is, however, an important, in fact basic,  difference between local and nonlocal theories. In the nonlocal case, despite the emergence of the familiar singularity structure, 
knowledge of this structure does not allow one to write dispersion relations of the standard type as discussed in section \ref{sing}. This is, of course, due to the fact that one is not able to close contours at infinity. This is ultimately connected to the subtle question of causality which we do not study in this paper.

As discussed in \cite{ELOP}, analyticity was adopted as a substitute for causality in the list of requirements on the physical S-matrix because of the difficulty of giving a sharp formulation of causality in terms of physical amplitudes. The connection between analyticity and causality is indeed somewhat murky already in local quantum field theory. The Bogoliubov causality condition \cite{BS} on amplitudes and the related well-known result in \cite{CN} do relate  singularity (pole) structure and causality. In any experiment, however, one always detects (on-shell) particles 
which imposes severe limitations on inferring corresponding space-time relations between interaction events.\footnote{This  is what makes the quantum problem qualitatively different from the corresponding classical field theory initial value problem \cite{T1}, where measurement of {\it fields} at a spacetime point  is assumed to be physically meaningful.} 
This is certainly made worse with nonlocal interactions where the Bogoliubov causality condition, or any other similar dispersion-related relations, cannot be derived in any analogous fashion. 
We hope to  address these questions elsewhere.

One of us (E. T. T.) would like to thank Ashoke Sen for discussions.

\setcounter{equation}{0}
\appendix
\renewcommand{\theequation}{\mbox{\Alph{section}.\arabic{equation}}}

\section{Appendix - Proofs of some statements in the main text} 

In this Appendix we provide the proofs of some assertions made in the main text. 

We first prove (\ref{Delta1}).  To this end we use the following known fact. Let $A$ be an $n\times n$ complex matrix, and $H(A) \equiv {1\over 2}( A + A^\dagger)$. Then if  $H(A)$ is a positive definite matrix, the following inequality holds: 
\beq
\det H(A) \leq |\det A|    \, . \label{det1}
\eeq
Equality holds iff $A= H(A)$, i.e., $A$ is hermitian. To prove (\ref{det1}) write $A= H(A) + S(A)$, where $S(A) = {1\over 2}( A- A^\dagger)$. Now, the asserted inequality (\ref{det1}) is the statement:  
\beq 
1 \leq  |\det [H(A)^{-1}  A] | = | \det [I + H(A)^{-1} S(A)] |   \, . 
\eeq
But 
\[ H(A)^{-1} S(A) = H(A)^{-1/2} \big[  H(A)^{-1/2} S(A)  H(A)^{-1/2}\big] H(A)^{1/2}  \, , \]
i.e., $ H(A)^{-1} S(A)$ is related by a similarity transformation to the matrix $[  H(A)^{-1/2} S(A)  H(A)^{-1/2}] $, which is anti-hermitian. Hence, it has purely imaginary eigenvalues $i\lambda_i$. But then, since for any real number $\lambda$ one has $| (1+ i\lambda )| = (1+ \lambda^2)^{1/2} \geq 1$, 
\beq 
  | \det [I + H(A)^{-1} S(A)] | = \prod_{i=1}^n | (1 + i\lambda_i)| \geq 1       \, , \label{det2}
  \eeq
which proves (\ref{det1}). Note that equality obtains if all $\lambda_i=0$, i.e., $S(A)=0$. 

We apply (\ref{det1}) to the matrix $d(w)$ given by (\ref{dmatrix}). One has 
\beq 
H(d(w))_{rs} = \sum_{i=1}^I \varepsilon^{(\ssc V)}_{ri}  \gamma_i( \varepsilon^{(\ssc V)})^ T_{is}  \, , \label{det3}
\eeq
where, with $\bar{\ell} > 0$, 
\beq
\gamma_i = \frac{\alpha_i + \bar{\ell}^{\,2}}{[ (\alpha_i + \bar{\ell}^{\, 2})^2 +\ell_2^4 \beta_i^2]}  > 0       \label{gamma}
\eeq
for all $\alpha \geq 0$, $-\infty < \beta_i < \infty$. 
$H(d(w))$ is manifestly positive definite, since 
\beq 
z^\dagger \!\cdot H(d(w))\!\cdot z = \sum_{i=1}^I \gamma_i\,  \Big|\sum_{s=1}^{V-1} \varepsilon^{(\ssc V)}_{si} z_s \Big|^2     > 0   \label{det4} 
\eeq
for any vector $z\in \mathbb{C}^{(V-1)}$. Applying (\ref{det1}) then: 
\beq 
| \det d(w) | > 0 \, .  \label{det5}
\eeq
It follows that 
\beq 
| \Delta(w)| = | \det d(w) |  \prod_{i=1}^I |w_i |   > 0   \, ,   \label{Delta1a}
\eeq
since $|w_i| = \big((\alpha_i + \bar{\ell}^2)^2 + \ell_2^4\beta_i^2\big)^{1/2} > 0$, all $i$.  

In the local case, where $\ell_1=\ell_2=0$, (\ref{Delta1a}) fails at $\alpha_i=0$. This is the source of UV divergences. 
The introduction of nonlocality then through either or both $\ell_i\not=0$ regulates all UV divergences as remarked in the main text. 

We next show the equivalence of the Landau equations (\ref{L2g}) - (\ref{L2h}) to their parametric form (\ref{L3d}) - (\ref{L3e}). 
Let $C_i$, $i=1, \cdots, L$, be a set of independent loops in the Feynman diagram $G$. The internal momenta $q_j$ may, if need be by a relabeling, be taken to be: $q_j= l_j$, for  $j=1, \cdots, L$, whereas each $q_j$ for  $j=L+1, \cdots, I$ is a linear combination of the loop momenta $l_i$ and the external momenta $P_r$ such that the momentum conservation system (\ref{momcon1}), (\ref{momcon2}) is satisfied. 

Introduce the vector $I$-component $K$ by 
\beq 
K_j = l_j \, , \quad j=1, \cdots, L\, ; \qquad K_{L+r} = P_r \, , \quad r=1, \cdots, (V-1)  \, .  \label{Kvec} 
\eeq
One can then write   
\beq 
q= R K   \, , \quad \mbox{or}  \quad  K= R^{-1} q  \, , \label{R1}
\eeq  
where the $(I\times I)$ matrix $R^{-1}$ is given by 
\begin{align}
R^{-1}_{ij} =& \delta_{ij} \, , \qquad i=1, \cdots, L ; \quad j=1, \cdots, I   \; , \nonumber \\ 
R^{-1}_{(L+r)j} =& \varepsilon_{rj} \, , \qquad  r=1, \cdots, (V-1) ; \quad   j=1, \cdots, I   \; ,  \label{R2}
\end{align} 
i.e., it has the structure 
\beq
R^{-1} = \left( \begin{array}{c} \begin{array}{lcr} 
{\bf 1}_{\ssc L}& \vdots    & {\bf 0}   \end{array}  \\
\cdots \cdots \cdots \\
\varepsilon^{(\ssc V)} 
\end{array} \right)   \, . \label{R3}
\eeq
Here, $\varepsilon^{(\ssc V)}$ denotes the incidence matrix with the $V$-th row deleted. 
Then, defining the quadratic form 
\beq 
{\cal Q}(l,P) \equiv \sum_{i=1}^I \alpha_i q_i^2   \, , \label{Qscr1} 
\eeq 
and introducing the $I\times I$ matrix $A$: 
\beq 
A_{ij}= \alpha_i \,\delta_{ij}  \, . \label{Amatrix}
\eeq
one has 
\beq 
{\cal Q}(l,P) = K^TR^T A RK   \, . \label{Qscr1a}  
\eeq

Now, given a quadratic form ${\cal Q}(z) = z^T M z$, where $M$ is an invertible matrix, the conjugate, or inverse,   quadratic form is defined to be $\tilde{\cal Q}(z) = z^T M^{-1}z$. 
If $z=(y,x)$, given the  quadratic form ${\cal Q}(z) = {\cal Q}(y,x)$ and its inverse $\tilde{\cal Q}(y,x)$, define 
$ {\cal Q}(x) \equiv {\rm extr}_{\D y} \, {\cal Q}(y,x)$.  

The following fact then holds \cite{S}: 
If $\tilde{\cal Q}(x)$ denotes the inverse of ${\cal Q}(x)$, one has 
\beq
 \tilde{\cal Q}(x) = \tilde{\cal Q}(0,x)   \, . \label{Sinvrel}
 \eeq
 Hence, ${\cal Q}(x)$ is given by the inverse of $\tilde{\cal Q}(0,x) $. 
 
Now, from (\ref{Qscr1a}) and using (\ref{R2}), (\ref{R3}), a simple computation gives 
\bea
\tilde{\cal Q}(l,P) & = & K^T R^{-1} A^{-1} (R^T)^{-1} K  \nonumber \\
  & = & \sum_{i=1}^L {1\over \alpha_i} \Big( l_i + \sum_{r=1}^{V-1} \varepsilon_{ri}P_r \Big)^2 + 
  \sum_{i=L+1}^I {1\over \alpha_i} \Big( \sum_{r=1}^{V-1} \varepsilon_{ri}P_r \Big)^2   \, . \label{Qscr2}
\eea
Hence, 
\bea 
\tilde{\cal Q}(0,P) & = & \sum_{i=1}^I {1\over \alpha_i} \Big( \sum_{r=1}^{V-1} \varepsilon_{ri}P_r \Big)^2  \nonumber \\
  & = & P d(\alpha) P   \,   . \label{Qscr3}
  \eea
Thus, 
\beq 
{\cal Q}(P) =  P d^{-1}(\alpha) P    \, , \label{Qscr4}  
\eeq
i.e., the extremum of (\ref{Qscr1}) over the loop momenta equals the quadratic form   $Q(\alpha; P)$  
in (\ref{S1}). Recalling the definition (\ref{psi}) of $\psi(\alpha, l, P)$ then, it immediately follows that equations (\ref{L3d}) - (\ref{L3e}) are equivalent to (\ref{L2g}) - (\ref{L2h}).

\end{document}